\documentclass[twocolumn,aps,floatfix,superscriptaddress]{revtex4}
\usepackage{array}
\usepackage{amsmath}
\usepackage{amssymb}
\usepackage{graphicx}
\usepackage{color}

\begin{document}

\title{Helical Antiferromagnetic Ordering in EuNi$_{1.95}$As$_2$ Single Crystals}

\author{N. S. Sangeetha}
\affiliation{Ames Laboratory, Iowa State University, Ames, Iowa 50011, USA}
\author{V. Smetana}
\affiliation{Department of Materials and Environmental Chemistry, Stockholm University, Svante Arrhenius v\"ag 16 C, 106 91 Stockholm, Sweden}
\author{A.-V. Mudring}
\affiliation{Department of Materials and Environmental Chemistry, Stockholm University, Svante Arrhenius v\"ag 16 C, 106 91 Stockholm, Sweden}
\author{D. C. Johnston}
\affiliation{Ames Laboratory, Iowa State University, Ames, Iowa 50011, USA}
\affiliation{Department of Physics and Astronomy, Iowa State University, Ames, Iowa 50011, USA}
\date{\today}

\begin{abstract}

The Eu$^{+2}$ spins-7/2 in EuNi$_2$As$_2$ with the body-centered tetragonal ${\rm ThCr_2Si_2}$ structure order antiferromagnetically below the N\'eel temperature $T_{\rm N} =15$~K into a helical antiferromagnetic (AFM) structure with the helix axis aligned along the tetragonal $c$~axis and the Eu ordered moments aligned ferromagnetically within the $ab$ plane as previously reported from neutron diffraction measurements [T.~Jin,~\emph{et al.}, Phys.\ Rev.\ B {\bf 99}, 014425 (2019)].  Here we study the crystallographic, magnetic, thermal, and electronic transport properties of Bi-flux-grown single crystals using single-crystal x-ray diffraction, anisotropic magnetic susceptibility $\chi$, isothermal magnetization $M$, heat capacity $C_{\rm_p}$, and electrical resistivity $\rho$ measurements versus applied magnetic field~$H$ and temperature~$T$\@. Vacancies are found on the Ni sites corresponding to the composition ${\rm EuNi_{1.95(1)}As_2}$.  A good fit of the $\rho(T)$ data by the Bloch-Gr\"uneisen theory for metals was obtained.  The $\chi_{ab}(T)$ data below $T_{\rm N}$ are fitted well by molecular field theory (MFT), and the helix turn angle $kd$ and the Eu-Eu Heisenberg exchange constants are extracted from the fit parameters. The $kd$ value is in good agreement with the neutron-diffraction result.  The magnetic contribution to the zero-field heat capacity below $T_{\rm N}$ is also fitted by MFT\@. The isothermal in-plane magnetization $M_{ab}$ exhibits two metamagnetic transitions versus~$H$, whereas $M_{c}(T = 2~{\rm K})$ is nearly linear up to $H=14$~T, both behaviors being consistent with MFT\@. The $M_c(H,T)$, $\rho(H_c,T)$, and $C_{\rm p}(H_c,T)$ data yielded a $H_c$-$T$ phase diagram separating the AFM and paramagnetic phases in good agreement with MFT\@.  Anisotropic $\chi(T)$ literature data for the ${\rm ThCr_2Si_2}$-type helical antiferromagnet ${\rm EuRh_2As_2}$ were also fitted well by MFT\@.  A comparison is made between the crystallographic and magnetic properties of ${\rm ThCr_2Si_2}$-type Eu$M_2Pn_2$ compounds with $M=$~Fe, Co, Ni, Cu, or Rh, and $Pn=$~P or~As, where only ferromagnetic and $c$-axis helical AFM structures are found.

\end{abstract}

\maketitle


\section{Introduction}

The body-centered tetragonal ${\rm ThCr_2Si_2}$ structure with space group $I4/mmm$~\cite{Just1996} is well known for accommodating  exotic ground state properties, rich magnetism, and  heavy fermion superconductivity~\cite{Stewart1984}. Compounds with this structure type also became extensively studied  after the discovery of superconductivity (SC) in the layered iron arsenides $A{\rm Fe_2As_2}$ (divalent $A=$ Ba, Sr, Ca, Eu) and related 122-type materials with SC transition temperatures up to $T_{\rm c}=38$~K~\cite{Johnston2010, Canfield2010, Stewart2011, Scalapino2012}.  The unconventional SC in doped $A{\rm Fe_2As_2}$ compounds arises adjacent in composition to a long-range ordered itinerant antiferromagnetic (AFM) spin-density-wave (SDW) phase associated with the Fe atoms, suggesting that magnetism and SC are closely intertwined in these systems~\cite{Johnston2010, Canfield2010, Stewart2011, Scalapino2012, Dai2012}.  

Compared to the ${\rm (Ca, Sr, Ba) Fe_2As_2}$-based compounds, the $A =$~Eu-based materials are different due to the Eu$^{+2}$ spin-7/2 $4f$ magnetic moments and associated rich magnetism. For example, ${\rm EuFe_2As_2}$ is a unique example in which localized ${\rm Eu^{+2}}$ spins order antiferromagnetically below 19~K with an $A$-type AFM structure and the itinerant Fe moments undergo an SDW transition at 190~K with an associated tetragonal-to-orthorhombic structural phase transition \cite{{Raffius1993},
{Ren2008},{Tegel2008},{Xiao2009}}. Similar to ${\rm (Ca, Sr, Ba) Fe_2As_2}$,  SC is achieved in  ${\rm EuFe_2As_2}$ by either chemical doping or by hydrostatic pressure after suppression of the SDW and the associated structural transition \cite{Terashima2009}.  The superconducting phase was found to coexist with long-range AFM from the Eu$^{+2}$ spin-7/2 sublattice, making the ${\rm EuFe_2As_2}$-based system very attractive for additional research \cite{Zapf2017}. The Eu spins in the isostructural compounds ${\rm EuCu_2P_2}$ and ${\rm EuRu_2As_2}$ also order magnetically \cite{{Huo2011},{Jiao2012}}.

The Eu- and Co-based pnictides ${\rm EuCo_2P_2}$ and ${\rm EuCo_2As_2}$ have also received considerable attention with respect to their magnetic properties. ${\rm EuCo_2P_2}$ has an uncollapsed tetragonal (ucT) structure and orders antiferromagnetically below 66~K with a coplanar helical magnetic structure at ambient pressure with no contribution from the Co atoms~\cite{{Morsen1988},{Reehuis1992}}. High-pressure studies on ${\rm EuCo_2P_2}$ showed that the system changes its magnetic character from Eu(4$f$)-sublattice ordering to Co(3$d$)-sublattice ordering coincident with a pressure-induced first-order ucT to collapsed tetragonal (cT) structural transition \cite{Chefki1998}. We recently showed that ${\rm EuCo_2P_2}$ at ambient pressure is a textbook example of a noncollinear helical AFM for which the thermodynamic properties in the AFM state are well described by our so-called unified molecular field theory (MFT) \cite{Sangeetha2016}. Similarly, single crystals of ${\rm EuCo_2As_2}$, which is isostructural and isoelectronic to ${\rm EuCo_2P_2}$, exhibit coplanar helical AFM ordering of ${\rm Eu^{+2}}$ spins below 47~K but with anomalously-enhanced effective and/or ordered moments \cite{{Raffius1993},{Marchand1978},{Tan2016},{Sangeetha2018}}. In contrast to ${\rm EuCo_2P_2}$, high-pressure studies on ${\rm EuCo_2As_2}$ showed a continuous ucT to cT crossover, which results in an intermediate-valence state of ${\rm Eu^{+2.25}}$ at high pressure \cite{Tan2016}.  Consequently, AFM ordering of the Eu sublattice gives way to ferromagnetic (FM) ordering  with a Curie temperature \mbox{$T\rm_C=$ 125~K} which arises from both Eu~4$f$ and Co~3$d$ moments \cite{{Bishop2010},{Tan2016}}. ${\rm EuCu_2As_2}$ \cite{Anand2015} and ${\rm EuRh_2As_2}$ \cite{Singh2009} also have the ${\rm ThCr_2Si_2}$-type structure and order antiferromagnetically at the N\'eel temperatures $T_{\rm N}=$17.5~K and~47~K, respectively. 

Superconductivity has been found in the Ni-based ${\rm ThCr_2Si_2}$-type compounds ${\rm SrNi_2As_2}$ \mbox{($T\rm_c=0.62$~K)}~\cite{Bauer2008}, ${\rm SrNi_2P_2}$ ($T\rm_c=1.4$~K)~\cite{Ronning2009}, ${\rm BaNi_2As_2}$ ($T\rm_c=0.7$~K)~\cite{Ronning2008}, and ${\rm BaNi_2P_2}$ \mbox{($T\rm_c=2.80$~K)}~\cite{Mine2008}. Moreover, it is well known that SC emerges in ${\rm EuFe_2As_2}$ when the long-range magnetic order is suppressed by hole doping in ${\rm Eu}_{1-x}{\rm K}_{1-x}{\rm Fe_2As_2}$~\cite{{Jeevan2008},{Maiwald2012}}, by electron doping with transition metal ions in ${\rm Eu [Fe}_{1-x}{\rm (Co, Ru, Ir)}_x{\rm ]_2As_2}$~\cite{{Jiao2012a},{Jiang2009},{Paramanik2013}}, or by isovalent substitution of As by P in ${\rm EuFe_2(As}_{1-x}{\rm P}_x{\rm )_2}$~\cite{{Ren2009},{Jeevan2011}}. However, Ni substitution on the Fe site in ${\rm EuFe_2As_2}$ does not induce superconductivity down to 2 K in contrast to Ni-doped ${\rm BaFe_2As_2}$~\cite{Li2009}. Instead, it was found that both the SDW transition and AFM ordering of ${\rm Eu^{+2}}$ moments were suppressed simultaneously by substituting Ni for Fe, and FM ordering of the ${\rm Eu^{+2}}$ moments emerges instead of SC~\cite{{Nowik2011},{Ren2009a}}.

The compound EuNi$_2$As$_2$ orders antiferromagnetically at $T_{\rm N}=15$~K~\cite{{Raffius1993},{Ghadraoui1988}}. Neutron-diffraction studies revealed that the Eu$^{+2}$ spins $S=7/2$ align ferromagnetically in the $ab$ plane and form an incommensurate AFM helical structure with the helix axis parallel to the tetragonal $c$~axis with magnetic propagation vector $k = [$0, 0, 0.9200(6)]$2\pi/c$~\cite{Jin2019}.  The ordered moments rotate in the $ab$~plane by $kc/2 =165.6(1)^\circ$ around the $c$ axis from layer to layer, indicating within molecular-field theory (MFT) that the dominant nearest-layer and next-nearest-layer interactions are both AFM\@. This study also showed that there is no contribution to the AFM ordering from the Ni sublattice.   

Herein we report a detailed study of EuNi$_2$As$_2$ single crystals including their crystallographic, magnetic, thermal, and electronic-transport properties, investigated using single-crystal x-ray diffraction (XRD), magnetic susceptibility $\chi(H, T)\equiv M(T)/H$, isothermal magnetization $M(H,T)$, heat capacity $C_{\rm p}(H,T)$, and electrical resistivity $\rho(H,T)$ measurements as functions of applied magnetic field~$H$ and temperature~$T$\@.

The experimental details are presented in Sec.~\ref{Sec:ExpDetails} and the crystallographic results in Sec.~\ref{Sec:crystaldata}.  A  new formulation of MFT was recently presented by one of us for calculating the magnetic and thermal properties of collinear and noncollinear AFMs on the same footing~\cite{Johnston2012, Johnston2015}, which was therefore dubbed the unified molecular-field theory.   This theory is applicable to systems of identical crystallographically-equivalent Heisenberg spins interacting by Heisenberg exchange and does not use the concept of magnetic sublattices. Instead, the magnetic properties are calculated solely from the exchange interactions of an arbitrary spin with its neighbors. In addition, the parameters of the MFT are experimentally measurable, replacing the vague molecular-field coupling constants of the traditional Weiss MFT\@.  The $M(H,T)$ isotherm and $\chi(H,T)$ data for single crystals are presented in Sec.~\ref{Sec:mag}, including analyses of these data by the MFT\@.  A good fit by the MFT to the anisotropic magnetic susceptibility of the helical Eu structure below~$T_{\rm N}$ was obtained for a helix turn angle in good agreement with the value found~\cite{Jin2019} from the neutron diffraction measurements.  In addition, the nearest- and next-nearest-interplane Heisenberg exchange interactions between the Eu spins in the $c$-axis helical structure were estimated from MFT analysis of the data.

The $ab$-plane $\rho(T)$ data are presented in Sec.~\ref{Sec:Res}, where an excellent fit by the Bloch-Gr\"uneisen theory was obtained.  Our $C_{\rm p}(H,T) $ data are presented in Sec.~\ref{Sec:HC}, where fits by MFT are presented.  The AFM-paramagnetic (PM) phase diagram in the $H_c-T$ plane was constructed from the $M(H_c,T)$, $C_{\rm p}(H_c,T)$ and $\rho(H_c,T)$ data.  A good fit by MFT to the boundary separating these two phases was obtained, yielding the extrapolated $c$-axis critical field $H_{c\perp}(T=0) = 13.6$~T\@.  A summary is given in Sec.~\ref{Sec:Summary}, which includes a comparison of the crystallographic and magnetic properties of ${\rm ThCr_2Si_2}$-type Eu$M_2Pn_2$ compounds with $M=$~Fe, Co, Ni, Cu, Rh and $Pn=$~P or~As.

\section{\label{Sec:ExpDetails} Experimental Details}

Single crystals of EuNi$_2$As$_2$ were  grown using both Bi flux and NiAs flux. For growths using NiAs flux, the starting materials were high-purity elemental Eu (Ames Laboratory), and Ni (99.999\%)  and As (99.99999\%) from Alfa Aesar. The EuNi$_2$As$_2$ and flux were taken in a 1:4  molar ratio and placed in an alumina crucible that was sealed under $\approx$ 1/4 atm high-purity argon in a silica tube. The sealed samples were preheated at 600~$^{\circ}$C for 5~h, and then heated to 1300~$^{\circ}$C at a rate of 50~$^{\circ}$C/h and held there for 15~h for homogenization. Then the furnace was slowly cooled at the rate of 6~$^{\circ}$C/h to 1180~$^{\circ}$C\@. The single crystals were separated by decanting the flux with a centrifuge at that temperature. Several 2$-$4~mm size shiny platelike single crystals were obtained from each growth.

Single crystals of EuNi$_2$As$_2$ were also grown in Bi flux with a purity of 99.999\% obtained from Alfa Aesar. EuNi$_2$As$_2$ and Bi were taken in a 1:10 molar ratio and placed in an alumina crucible that was sealed under argon in a silica tube. The sealed samples were preheated at  600~$^{\circ}$C for 6~h. Then the mixture were placed in a box furnace and heated to 1050~$^{\circ}$C at a rate of 50~$^{\circ}$C/h, held there for 20~h, and then cooled to 700~$^{\circ}$C at a rate of 2~$^{\circ}$C/h and then to 400~$^{\circ}$C at a rate of 5~$^{\circ}$C/h. At this temperature the molten Bi flux was decanted using a centrifuge. Shiny platelike crystals with basal-plane areas up to $2\times7$~mm$^2$ ($\approx 52$ mg) were obtained. 

Single-crystal XRD measurements were performed at room temperature on a Bruker D8 Venture diffractometer operating at 50~kV and 1~mA equipped with a Photon 100 CMOS detector, a flat graphite monochromator, and a Mo~K$\alpha$ I$\mu$S microfocus source ($\lambda = 0.71073$~\AA). The preliminary quality testing was performed on a set of 32 frames. The raw frame data were collected using the Bruker APEX3 software package \cite{APEX2015}.  The frames were integrated with the Bruker SAINT program~\cite{SAINT2015} using a narrow-frame algorithm integration  and the data were corrected for absorption effects using the multi-scan method (SADABS) \cite{Krause2015}. The occupancies of the atomic positions were refined assuming random occupancy of the Ni and As sites and complete occupancy of the Eu sites.  The atomic displacement parameters were refined anisotropically.  Initial models of the crystal structures were first obtained with the program SHELXT-2014 \cite{Sheldrick2015A} and refined using the program SHELXL-2014 \cite{Sheldrick2015C} within the APEX3 software package.

The phase purity and chemical composition of the EuNi$_2$As$_2$ crystals were studied using an energy dispersive x-ray spectroscopy (EDS) semiquantitative chemical analysis attachment to a JEOL scanning electron microscope (SEM).  SEM scans were taken on cleaved surfaces of the crystals which verified the single-phase nature of the crystals. The composition of each platelike crystal studied here was measured at six or seven positions on each of the two basal $ab$-plane faces, and the results were averaged.  Good chemical homogeneity was found for each crystal.  The chemical compositions of EuNi$_2$As$_2$ crystals obtained from both the EDS and single-crystal x-ray structural analysis were determined  assuming that the Eu site is fully occupied. The same crystals measured by EDS (pieces of which were used for the XRD measurements) were utilized to perform the physical-property measurements.

Magnetization data were obtained using a Quantum Design, Inc., SQUID-based magnetic-properties measurement system (MPMS) in magnetic fields up to 5.5~T and a vibrating-sample magnetometer (VSM) in a Quantum Design, Inc., physical-properties measurement system (PPMS) in magnetic fields up to 14~T where 1~T~$\equiv10^4$~Oe. The magnetic moment output of these instruments is expressed in Gaussian cgs electromagnetic units (emu), where 1~emu = 1~G\,cm$^3$ and 1~G = 1~Oe.  The $C_{\rm p}(H,T)$ was measured by a relaxation technique using a PPMS\@. The $\rho(H,T)$ measurements were performed using a standard four-probe ac technique using the ac-transport option of the PPMS with the current in the $ab$~plane. Annealed platinum wire (25 $\mu$m diameter) electrical leads were attached to the crystals using silver epoxy.

\section{\label{Sec:crystaldata} Crystallography}

\begin{table*}
\caption{\label{Table:SCXRD} Crystal and refinement parameters and atomic coordinates obtained from Rietveld refinement of room-temperature single-crystal XRD data of  EuNi$_2$As$_2$  with the ${\rm ThCr_2Si_2}$-type crystal structure, space group $I4/mmm$ and $Z=2$ formula units per unit cell.  Compositions obtained from energy-dispersive x-ray spectroscopy (EDS) measurements are also shown.}
\begin{ruledtabular}
\begin{tabular}{cc|cc| cc| ccc}
\multicolumn{2}{c}{Composition} &&& \multicolumn{2}{c}{Occupancy (\%)}   \\

XRD								&	EDS									&Atom 				& Wyckoff		&XRD	&EDS		& $x$			& $y$  	& $z$		\\
									&										&					& position		 &		&			&				&		&\\

\hline 

${\rm EuNi_{1.95(1)}As_{1.98(1)}}$\footnotemark[1]	&${\rm EuNi_{1.98(4)}As_{1.95(10)}}$\footnotemark[1]		&Eu  			& 2$a$  		&100		&100	& 0  		& 0  		& 0			\\
									&										&Ni			& 4$d$  		&97.7(6)	&99.0(2)		& 0  		& 1/2  	& 1/4		\\				
									&										&As  		& 4$e$ 		 &99.0(4)	&97(5)			& 0  		& 0  		& 0.36639(9)	\\
									&										&						&			 &		&			&				&		&\\

${\rm EuNi_{1.87(1)}As_2}$\footnotemark[2]	&${\rm EuNi_{1.84(1)}As_{2.01(4)}}$\footnotemark[2]		&Eu  					& 2$a$  		&100		&100		& 0  		& 0  		& 0			\\
									&										&Ni			& 4$d$ 		 &93.4(4)	&92.2(5)	& 0  		& 1/2  	& 1/4		\\
									&										&As  		& 4$e$  		&100.0(8)	&105(2)	& 0  		& 0  		& 0.36653(8)	\\	
&& &   &  &  & &&\\
${\rm EuNi_2As_2}$\footnotemark[3]	(Ref.~\cite{Jeitschko1988})												&&Eu  					& 2$a$ 		 &100	&		& 0  		& 0  		& 0			\\
																			&&Ni  					& 4$d$  		& 94(4)	&		& 0  		& 1/2  	& 1/4		\\
																			&&As					& 4$e$  		& 96(3)	&		& 0  		& 0  		& 0.3669(4)	\\
\hline 
\hline
Lattice parameters&  & EuNi$_2$As$_2$\footnotemark[1]   &  & EuNi$_2$As$_2$\footnotemark[2]   & Ref.~\cite{Jeitschko1988}\footnotemark[3] \\
\hline
				
$a$~(\AA)					&& 4.105(2)  & &4.1052(8) 	&4.115(1)	\\
$c$~(\AA)					&& 10.078(4) &&10.027(2) 	&10.086(2)	\\
$c/a$  					&& 2.455(2) && 2.442(1)		&2.451(1)		\\
$V_{\rm cell}~(\rm{\AA}^3)$	&& 169.8(2) &&168.99(7)		&170.78(11)	\\
\end{tabular}
\end{ruledtabular}
\footnotetext[1]{Grown in Bi flux}
\footnotetext[2]{Grown in NiAs flux}
\footnotetext[3]{Polycrystalline sample}
\end{table*}

The chemical compositions and crystal data for the EuNi$_2$As$_2$ single crystals grown in both Bi flux and NiAs flux obtained from the single-crystal XRD and EDS measurements at room temperature are presented in Table~\ref{Table:SCXRD}. The data confirm that EuNi$_2$As$_2$ has the ${\rm ThCr_2Si_2}$-type body-centered tetragonal symmetry with space group $I4/mmm$, consistent with previous reports \cite{Jeitschko1988}. However, we consistently found randomly-distributed vacancies on the Ni site from both EDS and single-crystal XRD measurements as shown in Table~\ref{Table:SCXRD}, indicating that the composition of the crystals is EuNi$_{1.95(1)}$As$_2$. We find no significant difference in the lattice parameters between Bi-flux-grown and NiAs-flux-grown crystals. Therefore, all the physical-property measurements reported below were performed on the Bi-flux-grown crystals because large high-quality homogeneous single crystals of EuNi$_2$As$_2$ could be more easily grown with this flux.

\section{\label{Sec:mag} Magnetization and magnetic susceptibility measurements}

\subsection{Magnetic Susceptibility Measurements}

\begin{figure}
\includegraphics[width=3in]{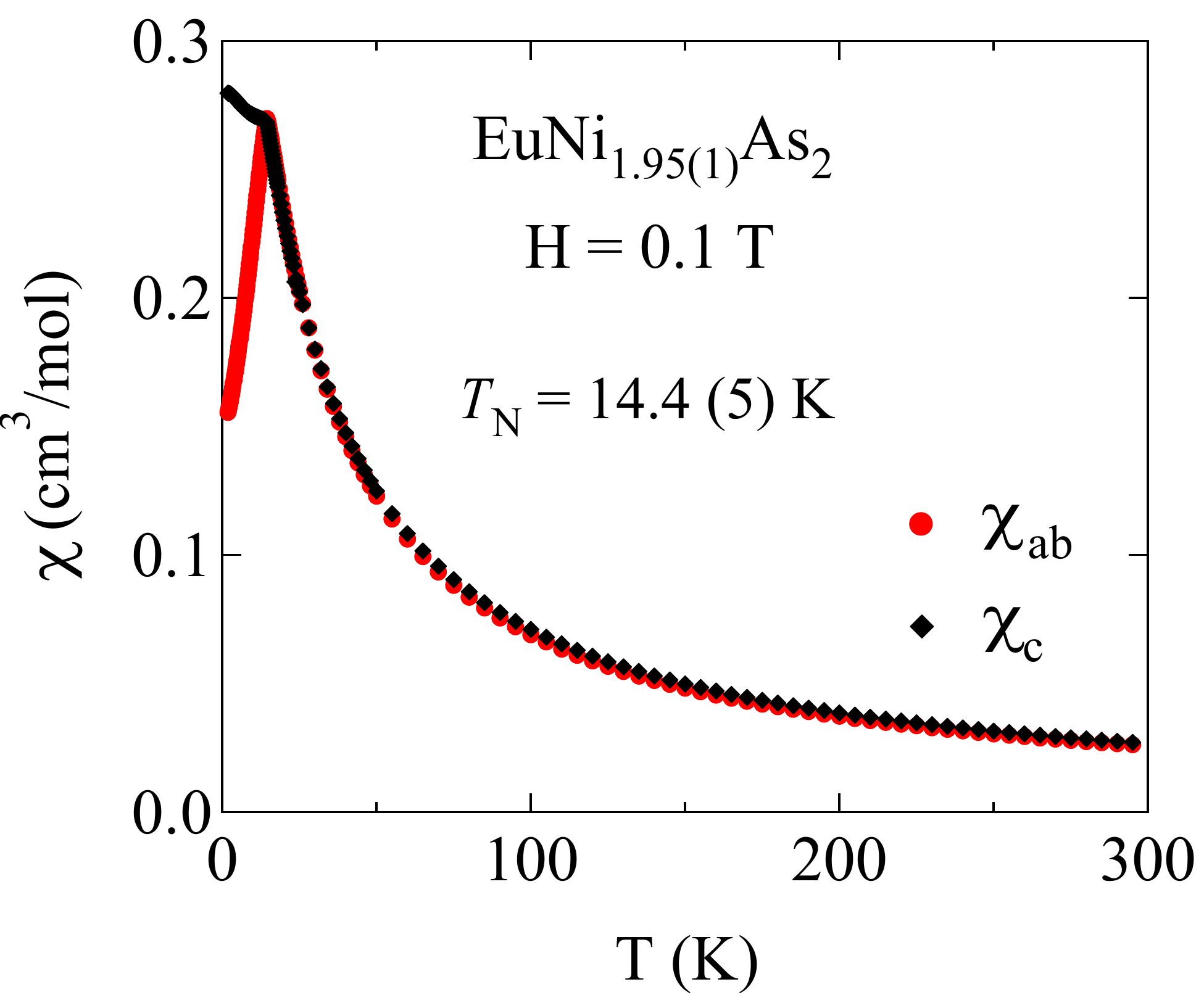}
\caption{Zero-field-cooled (ZFC) magnetic susceptibility \mbox{$\chi(T)\equiv M(T)/H$} of an ${\rm EuNi_{1.95}As_2}$ single crystal as a function of temperature~$T$ between 1.8 to 300~K measured in magnetic field $H=0.1$ T applied in the $ab$~plane ($\chi_{ab}$) and along the $c$~axis ($\chi_c$).}
\label{Fig:chi}
\end{figure}

The zero-field-cooled (ZFC) magnetic susceptibilities $\chi(H,T) \equiv M(T)/H$ of an EuNi$_{1.95}$As$_2$ single crystal measured in $H=0.1$~T aligned along the $c$~axis \mbox{($\chi_c,~H\parallel c$)} and in the $ab$~plane ($\chi_{ab}, ~H\parallel ab$) are shown in Fig.~\ref{Fig:chi}. A sharp peak in $\chi_{ab}(T)$ occurs at $T_{\rm N} = 14.4(5)$~K, in good agreement with the previous reports~\cite{{Raffius1993},{Ghadraoui1988},{Jin2019}}. The anisotropic $\chi(T)$ data below $T_{\rm N}$, where $\chi_c$ is nearly independent of~$T$ and $\chi_{ab}$ decreases with decreasing~$T$, indicate that the AFM-ordered moments are aligned in the $ab$ plane. Moreover, based on MFT, the nonzero limit of $\chi_{ab}(T\rightarrow 0)$ indicates that the AFM ordering in EuNi$_2$As$_2$ is either a collinear AFM with multiple domains in the $ab$~plane or an intrinsic coplanar noncollinear AFM structure~\cite{Johnston2012, Johnston2015}. The recent neutron-diffraction study of single-crystal EuNi$_2$As$_2$ indeed showed an incommensurate AFM helical structure with the Eu ordered moments aligned ferromagnetically within the $ab$~plane which rotate about the $c$~axis by 165.6(1)$^\circ$ from Eu~layer to Eu~layer along the $c$~axis, where the AFM propagation vector is \mbox{$k =(0,0,0.9200)2\pi/c$} and $c$ is the tetragonal $c$-axis lattice parameter~\cite{Jin2019}. Within MFT, this turn angle indicates dominant AFM nearest-layer and also next-nearest-layer Eu-Eu interactions~\cite{Johnston2012, Johnston2015}.  Similar incommensurate Eu helical spin structures along the $c$~axis with almost the same propagation vector were found in the isostructural compounds ${\rm EuCo_2P_2}$ and ${\rm EuCo_2As_2}$~\cite{Reehuis1992, Sangeetha2016, Tan2016, Sangeetha2018}. 

\begin{table*}
\caption{\label{Tab.chidata} Parameters obtained from Curie-Weiss and modified Curie-Weiss fits to $\chi(T)$ data for $H=0.1$~T and \mbox{$H=1$~T} between 70 and 300~K for EuNi$_{1.95}$As$_2$ using Eqs.~(\ref{Eq.cw}) and~(\ref{Eq.mcw}), respectively. Listed are the temperature-independent contribution $\chi_{0\alpha}$  with the field applied in the $\alpha = ab, c$ directions, molar Curie constant $C_\alpha$, angle-averaged molar Curie constant $C_{\rm {ave}}=(2 C_{ab} +C_c)/3$, effective moment $\mu_{\rm eff}$, angle-averaged effective moment $\mu_{\rm eff,ave}$, Weiss temperature $\theta\rm_{p\alpha}$, angle-averaged Weiss temperature $\theta\rm_{p,ave}$, and $f_\alpha\equiv\theta_{\rm p\alpha}/T_{\rm N}$ with $T_{\rm N} = 14.5$~K\@. The error bars reflect systematic errors found from different temperature ranges of the fits. The effective moment per Eu atom $\mu{\rm_{eff,ave}(\mu_B/f.u.)}$ was calculated from Eq.~(\ref{Eq:mueff}). For $S=7/2$ with $g=2$, Eqs.~(\ref{Eq:cc}) and (\ref{Eq:mueff}) yield $C=7.878~{\rm cm^3\,K/(mol~Eu)}$ and $\mu_{\rm eff} = 7.937~\mu{\rm_B/Eu}$, respectively.}
\begin{ruledtabular}
\begin{tabular}{ccccccccccc}	
	
Field	&
  	&$\chi_{0\alpha}$		
	&  $C_{\alpha}$ 	
	&  $C_{\rm ave}$ 	
	&  $\mu_{\rm eff\alpha}$ 	 
	& $\mu_{\rm\rm eff, ave}$       
	& $\theta_{\rm p\alpha}$ 
	&$\theta\rm_{p,ave}$
	&$\theta\rm_{p,diff}$
	& $f_\alpha\equiv\frac{\theta_{\rm p\alpha}}{T_{\rm N}}$	\\
	&&$\rm{\left(10^{-4}\frac{cm^3}{mol}\right)}$	
	&  $\rm{\left(\frac{cm^3 K}{mol}\right)}$ 
 	&  $\rm{\left(\frac{cm^3 K}{mol}\right)}$ 
	 & $\rm{\left(\frac{\mu_B}{Eu}\right)}$ 
	& $\rm{\left(\frac{\mu_B}{Eu}\right)}$   
	& (K) 	
	& (K)		
	& (K)		
 	&	\\
\hline
$H=0.1$~T& $H\parallel ab$ 		& 10.1(2)		& 7.8(1) 		&7.91(7)	&7.90(5)	&7.95(3)	&$-15(1)$	&$-15.2(8)$	& 0.6	& 	$-1.03$ \\
& $H\parallel c$ 		&10.9(8)		&8.13(3) 		&		&8.06(1) 	&		&$-15.6(5)$	&			&		&	$-1.07$\\
& $H\parallel ab$ 		&$\equiv 0$	&8.24(4) 		&8.3(3)	&8.1(2)	&8.2(1)	&$-19(1)$	&$-19.4(9)$	& 1.2	&	$-1.31$ \\
& $H\parallel c$ 		&$\equiv 0$	&8.57(3) 		&		&8.28(1) 	&		&$-20.2(9)$	&			&		&	$-1.39$\\
\hline
$H=1$~T& $H\parallel ab$ 		& 6.2(2)		& 7.68(1) 		&7.75(1)	&7.838(5)	&7.874(5)	&$-11.7(1)$	&$-12.7(1)$	& $-3.0$	& 	$-0.81$ \\
& $H\parallel c$ 		&4.4(2)		&7.90(1) 		&		&7.949(5) 	&		&$-14.7(1)$	&			&		&	$-1.01$\\
& $H\parallel ab$ 		&$\equiv 0$	&7.938(5) 		&7.975(5)	&7.969(2)	&7.987(2)	&$-14.9(1)$	&$-15.63(1)$	& $-2.2$	&	$-1.03$ \\
& $H\parallel c$ 		&$\equiv 0$	&8.097(2) 		&		&8.048(1) 	&		&$-17.1(1)$	&			&		&	$-1.18$\\

\end{tabular}
\end{ruledtabular}
\end{table*}

\begin{figure}
\includegraphics[width=2.8in]{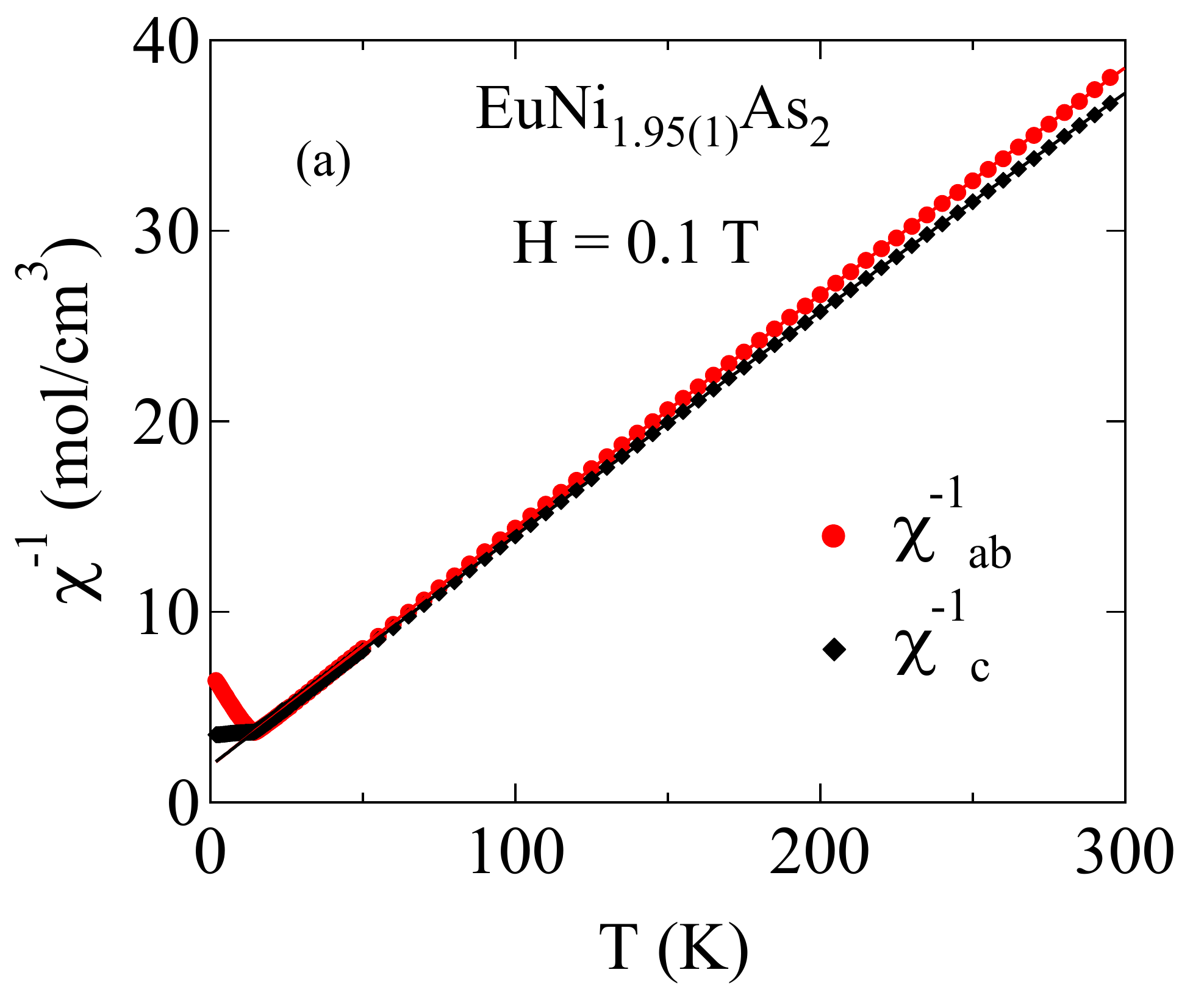}
\includegraphics[width=2.8in]{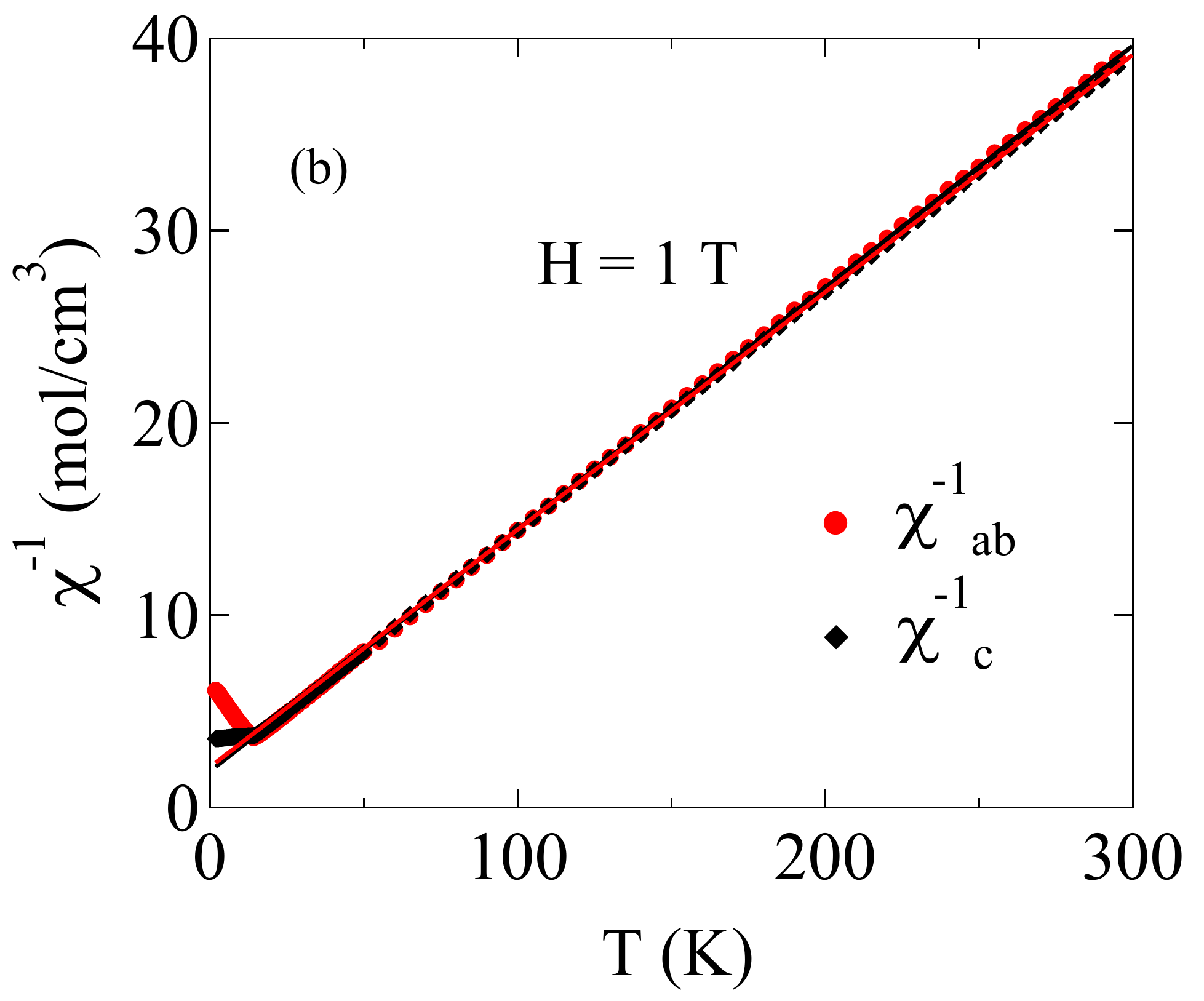}
\caption{Inverse susceptibility $\chi^{-1}$ versus temperature $T$ of an ${\rm EuNi_{1.95}As_2}$ single crystal for (a)~$H=0.1$~T and (b)~$H=1$~T applied in the $ab$~plane ($\chi^{-1}_{ab}$, $H \parallel ab$) and along the $c$~axis ($\chi^{-1}_c$, $H \parallel c$).}
\label{Fig:invchi}
\end{figure}

The inverse susceptibilities $\chi_{ab}^{-1}(T)$ and $\chi_c^{-1}(T)$ of EuNi$_{1.95}$As$_2$ measured in $H=0.1$~T and $H=1$~T are shown in Figs.~\ref{Fig:invchi}(a) and ~\ref{Fig:invchi}(b), respectively. The high-temperature $\chi^{-1}(T)$ data above $T_{\rm N}$ in the PM region 70~K~$\leq T \leq 300$~K were fitted by both the Curie-Weiss law and the modified-Curie-Weiss law, given respectively by  
\begin{subequations}
\label{Eq.chi}
\begin{equation}
\chi_{\alpha}(T) = \frac{C_{\alpha}}{T-\theta_{\rm p\alpha}}
\label{Eq.cw}
\end{equation}
and
\begin{equation}
\chi_{\alpha}(T) =\chi_{0\alpha} + \frac{C_{\alpha}}{T-\theta_{\rm p\alpha}},
\label{Eq.mcw}
\end{equation}
\end{subequations}
where $\alpha=ab$ or~$c$ and $\chi_{0\alpha}$ is a \mbox{$T$-independent} term. The Curie constant per mole of Eu spins is given by
\begin{subequations}
\begin{equation}
C_{\alpha}=\frac{N_{\rm A} g^2_{\alpha}S(S+1)\mu^2_{\rm B}}{3k_{\rm B}},
\label{Eq:cc}
\end{equation}
where $N_{\rm A}$ is Avogadro's number, $g_\alpha$ is the possibly anisotropic spectroscopic splitting factor ($g$-factor), and $k_{\rm B}$ is Boltzmann's constant.  The effective moment $\mu_{{\rm eff}\alpha} = g\sqrt{S(S+1)}$ of a spin in units of $\mu_{\rm B}$ is given by Eq.~(\ref{Eq:cc}) as
\begin{equation}
\mu_{\rm eff \alpha} = \sqrt{\frac{3k_{\rm B}C_{\alpha}}{N_{\rm A}{\rm\mu^2_B} }} ~ \approx \sqrt{8C_{\alpha}},
\label{Eq:mueff}
\end{equation}
\end{subequations}
where $C_\alpha$ is in cgs units of cm$^3$\,K/(mol Eu$^{+2}$). The fits are shown as the straight lines in Fig.~\ref{Fig:invchi} and the fitted parameters together with the parameter $f_\alpha\equiv\theta_{\rm p\alpha}/T_{\rm N}$ used later are listed in Table~\ref{Tab.chidata}.

The values of $C\rm_{ave}$ and $\mu\rm_{eff, ave}$ in Table~\ref{Tab.chidata} are similar to the theoretical values of $7.878~{\rm cm^3\,K/mol~Eu}$ and $7.937~\mu{\rm_B/Eu}$, respectively, for ${\rm Eu^{+2}}$ spins with $S=7/2$ and $g=2$.  This suggests that the Ni atoms are nonmagnetic, as also inferred from the neutron-diffraction study of the magnetic structure~\cite{Jin2019}.  In the latter study the effective moment in the PM state was found to be $\mu\rm_{eff}=8.23(5)~\mu_B$/Eu from $\chi_{ab}(T)$ in $H=0.1$~T, which agrees with our value for $H=0.1$~T and $\chi_0=0$ listed in Table~\ref{Tab.chidata}.  However, when $\chi_0$ is a fitted parameter, we obtain the value $\mu\rm_{eff, ave} = 7.91(7)~\mu_{\rm B}$/Eu, close to that expected for Eu$^{+2}$ with $S=7/2$ and $g=2$.

Our value $\theta_{{\rm p}, ab}= -17(2)$~K of the Weiss temperature with $H\parallel ab$ in Table~\ref{Tab.chidata} for $H=0.1$~T is the same within the errors as that [$-17.7(9)$~K] obtained for a crystal with the same field and field orientation in Ref.~\cite{Jin2019}, indicating predominantly AFM exchange interactions between the Eu spins. The difference in the Weiss temperature between the $c$-axis and $ab$-plane values for $H=0.1$~T in Table~\ref{Tab.chidata} is small ($\lesssim 1$~K), indicating a small anisotropy field.  That this anisotropy field is small is consistent with the occurrence of two metamagnetic transitions in $M_{ab}(H)$ isotherms in Fig.~\ref{Fig:MH2K} below, as explained at the end of the following section.  The slight anisotropy may arise from either the anisotropic magnetic dipole interactions between the Eu spins~\cite{Johnston2016}, from single-ion uniaxial anisotropy~\cite{Johnston2017}, and/or from anisotropy in the Ruderman-Kittel-Kasuya-Yosida (RKKY) interactions between the Eu spins.

\begin{figure}
\includegraphics[width=2.8in]{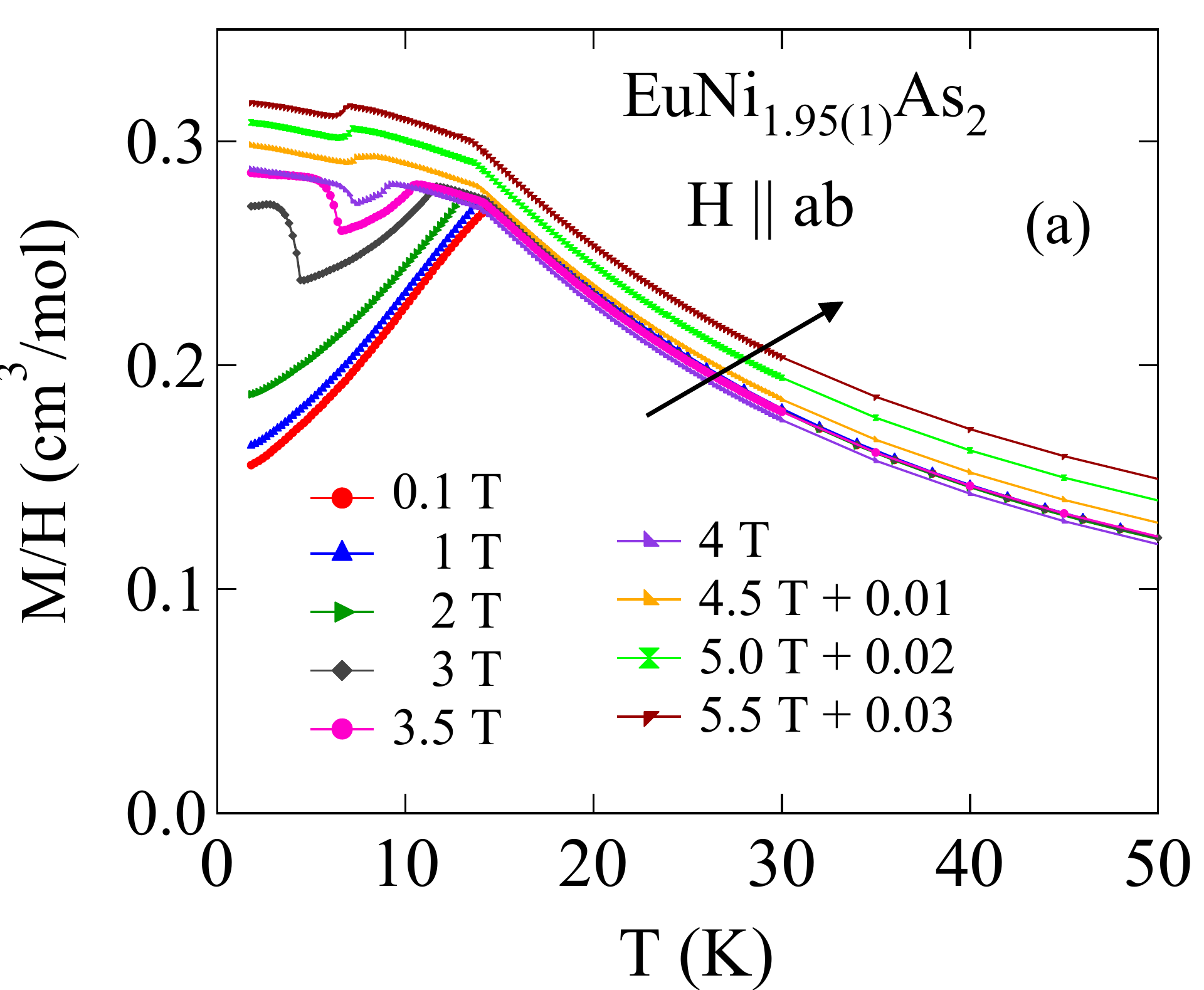}
\includegraphics[width=2.8in]{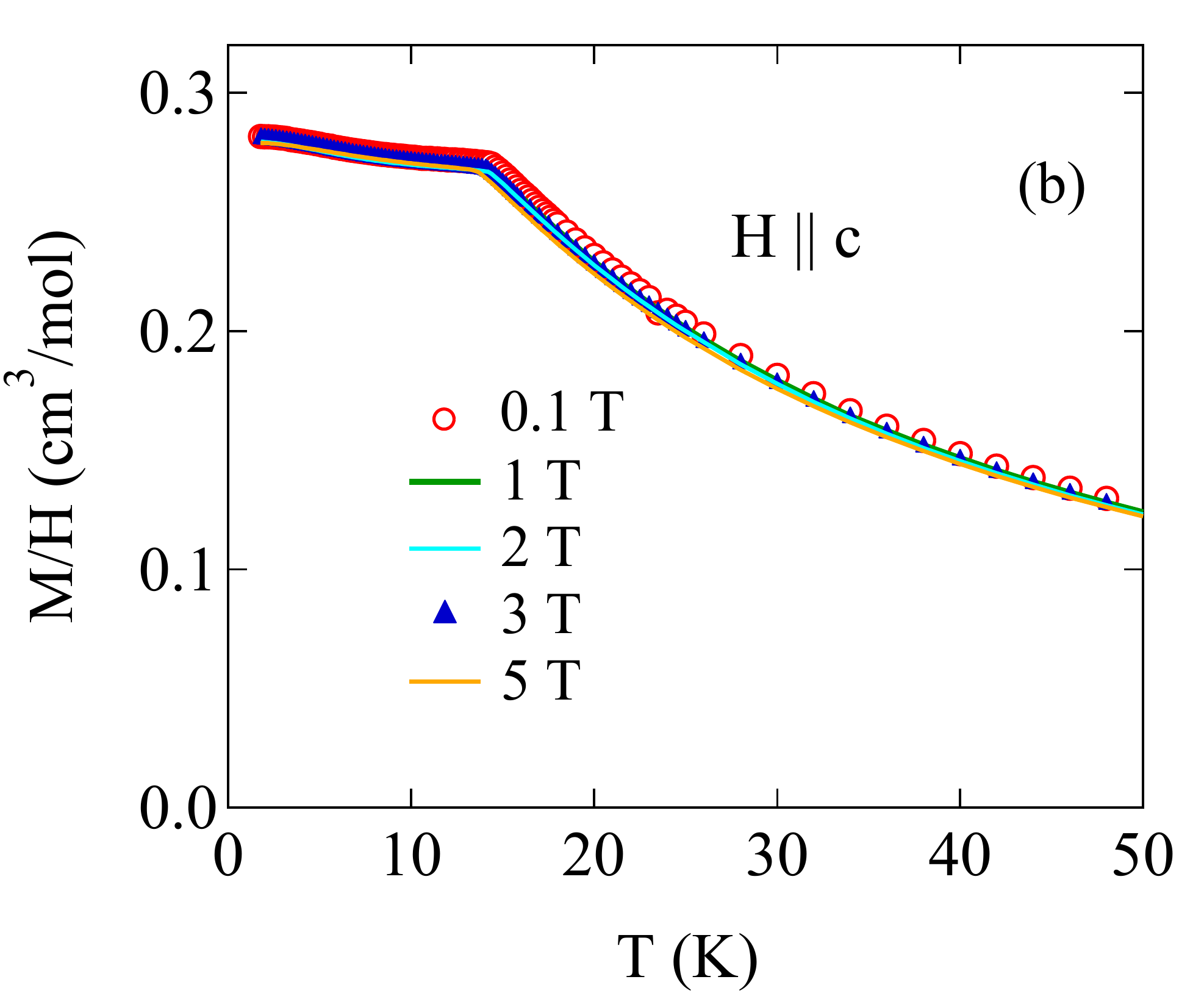}
\caption{Zero-field-cooled field-dependent magnetic susceptibility $\chi\equiv M/H$ of ${\rm EuNi_{1.95}As_2}$ single crystal as a function of temperature~$T$ for various magnetic fields $H$ applied (a) in the $ab$~plane ($H \parallel ab$) and (b) along the $c$~axis ($H \parallel c$). The data for $H\parallel ab=4.5,~5$ and 5.5~T are offset from each other for clarity by 0.01 cm$^3$/mol as indicated in~(a).}
\label{Fig:chifull}
\end{figure}

The field-dependent ZFC magnetic susceptibilities $\chi_\alpha\equiv M_\alpha(T,H)/H$ measured at various magnetic fields applied in the $ab$~plane and along the $c$~axis for the temperature range \mbox{$1.8~{\rm K}\leq T \leq 50~{\rm K}$} are shown in Figs.~\ref{Fig:chifull}(a) and \ref{Fig:chifull}(b), respectively.  Metamagnetic transitions are seen to occur in $\chi_{ab}(H,T)$ for $H\geq3$~T, as might be expected from the helical magnetic structure~\cite{Johnston2017b, Johnston2019}. One sees that $\chi_c(T)$ in Fig.~\ref{Fig:chifull}(b) is far less sensitive to $H$ compared to $\chi_{ab}(T)$ in Fig.~\ref{Fig:chifull}(a).  

\subsection{Magnetization versus Field Isotherms}
 
\begin{figure}
\includegraphics[width=2.8in]{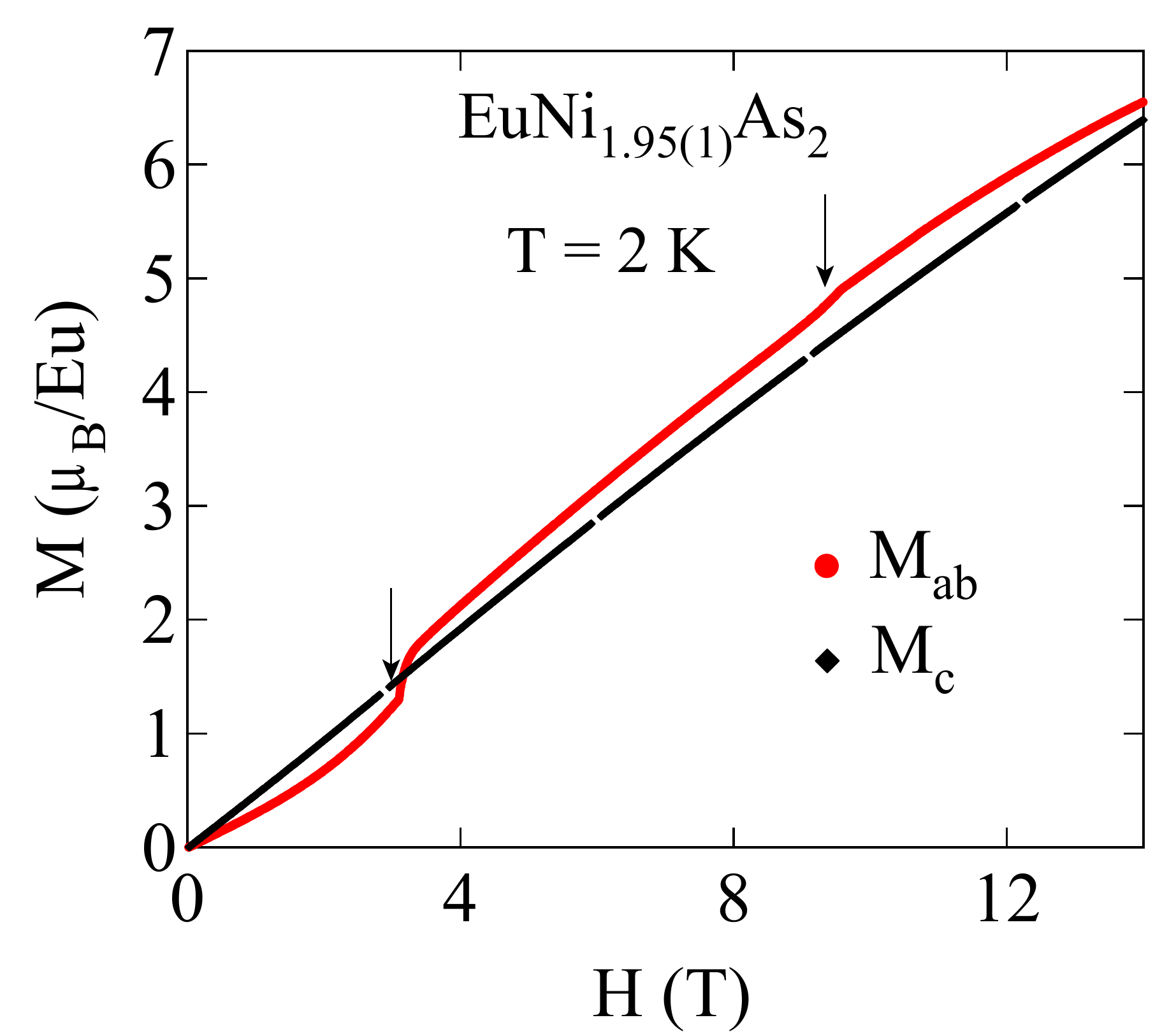}
\caption{Isothermal magnetization $M$ of single-crystal ${\rm EuNi_{1.95}As_2}$ as a function of magnetic field $H$ measured at $T=2$~K with $H$ applied in the $ab$~plane ($M_{ab}, H\parallel ab$) and along the $c$~axis ($M_{c}, H\parallel c$).  The first-order and second-order transition fields with increasing field for $M_{ab}(H)$ are indicated by vertical arrows, respectively.}
\label{Fig:MH2K}
\end{figure}

Isothermal $M(H)$ data measured at $T=2$~K with \mbox{$0\leq H \leq 14$~T} applied in the $ab$ plane ($M_{ab}, H\parallel ab$) and along the $c$-axis ($M_{c}, H\parallel c$) are shown in Fig.~\ref{Fig:MH2K}. The $M_c(H)$ data  are nearly linear in field at $T\ll T_{\rm N}$ as predicted by MFT for a helix with the field applied along the helix axis \cite{Johnston2015}. One sees that $M_c(H=14~{\rm T}) = 6.39~\mu\rm_B/Eu$ does not yet reach the saturation moment per Eu spin given by $\mu{\rm_{sat}}=gS\mu\rm_B=7~\mu\rm_B$ for $S=7/2$ and $g=2$. 

For $H\parallel ab$, the $M_{ab}(H)$ data in Fig.~\ref{Fig:MH2K} show an apparently first-order metamagnetic transition at a field $H\rm_{mm1} \approx 3.1$~T (marked by a vertical arrow). Then at a higher field $H_{\rm mm2} \approx 9.5$~T another metamagnetic transition occurs, marked by another vertical arrow, which appears to be second order.

\begin{figure}
\includegraphics[width=2.8in]{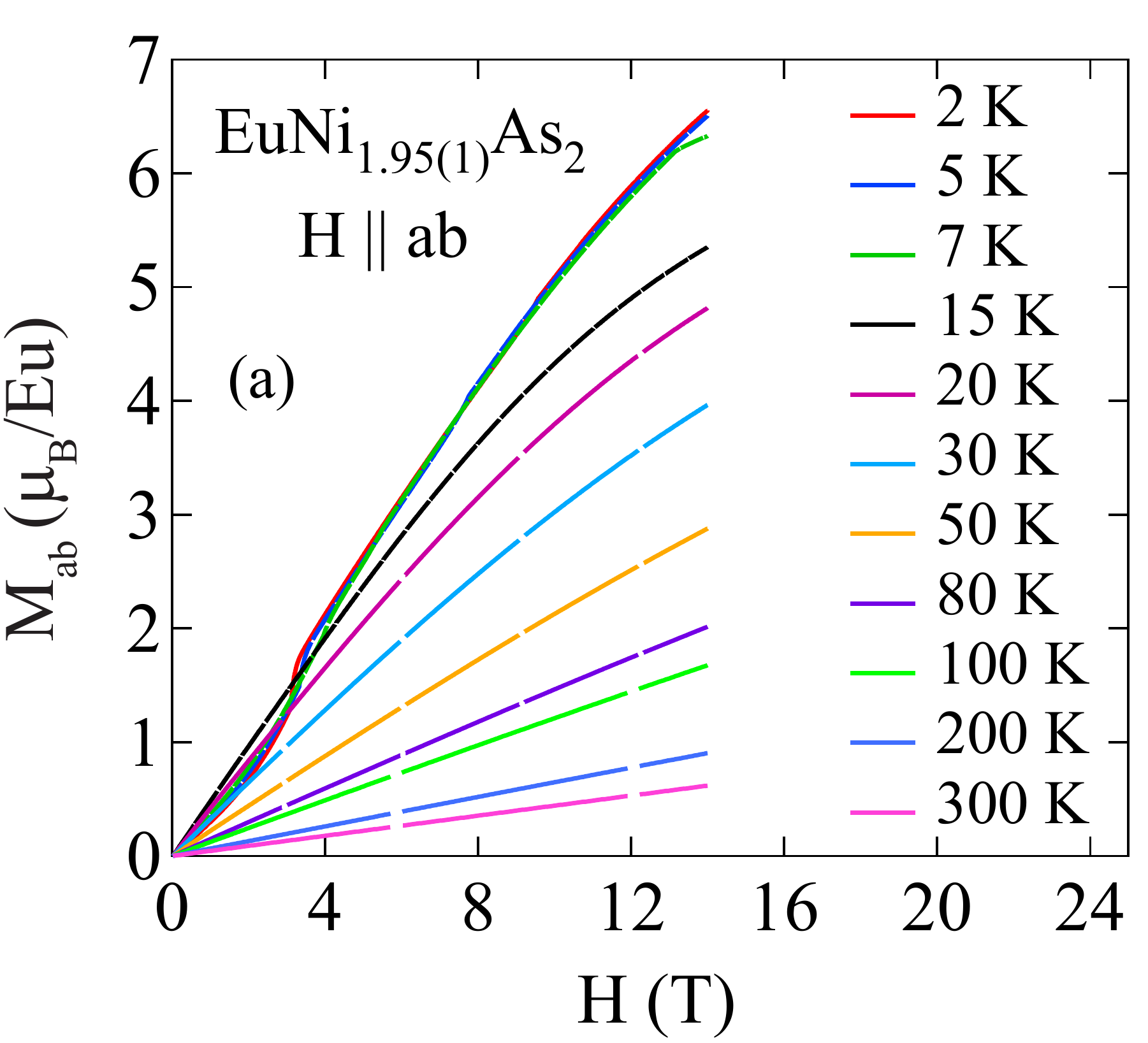}
\includegraphics[width=2.8in]{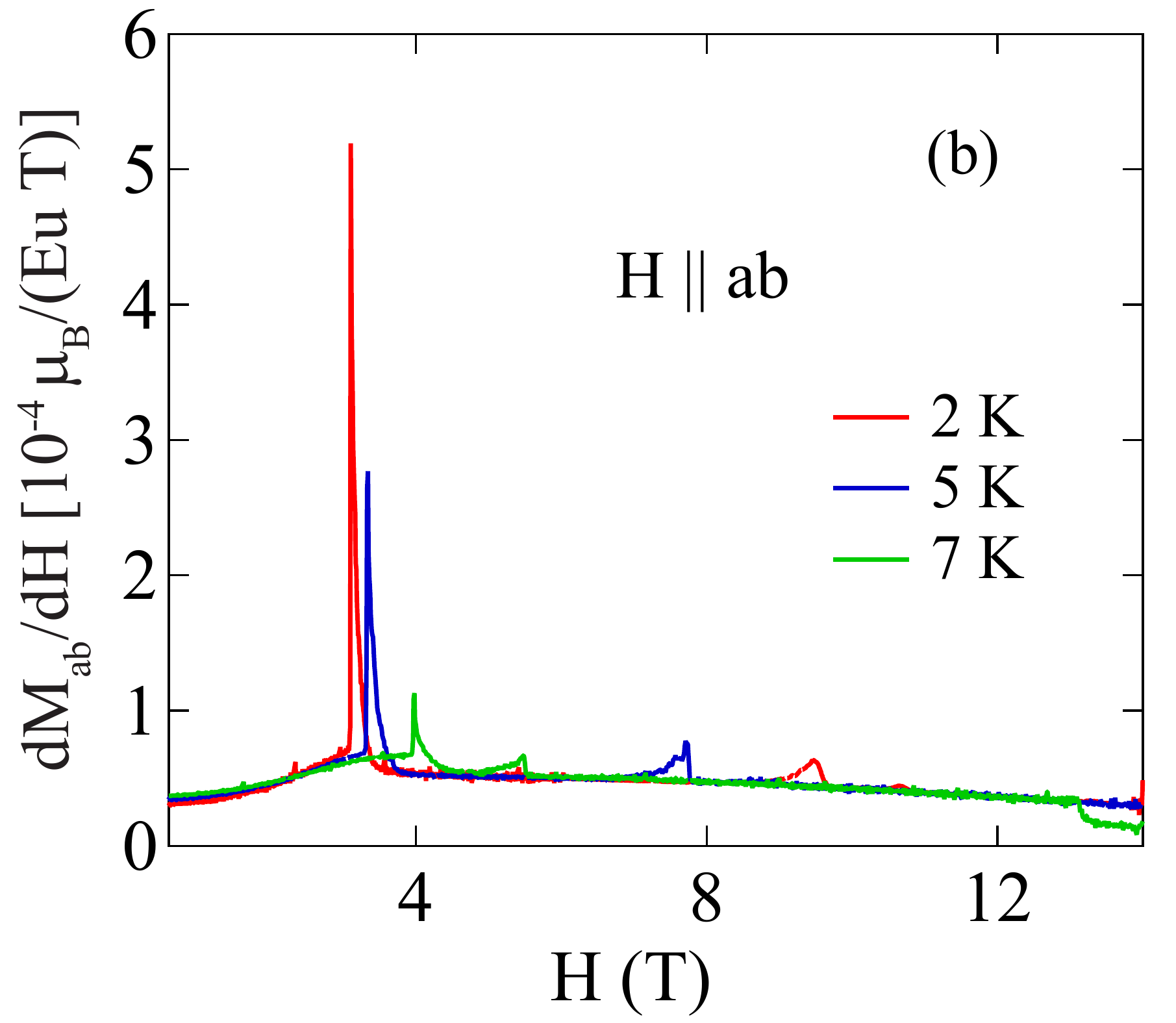}
\caption{(a) Isothermal magnetization $M_{ab}$ of an ${\rm EuNi_{1.95}As_2}$ single crystal versus magnetic field~$H$ applied in the $ab$ plane ($H\parallel ab$) at the indicated temperatures~$T$\@. (b)~Derivative $dM_{ab}/dH$  versus~$H$ at $T = 2,$ 5, and 7~K obtained from the respective data in (a). }
\label{Fig:MHab}
\end{figure}

\begin{figure}
\includegraphics[width=2.6in]{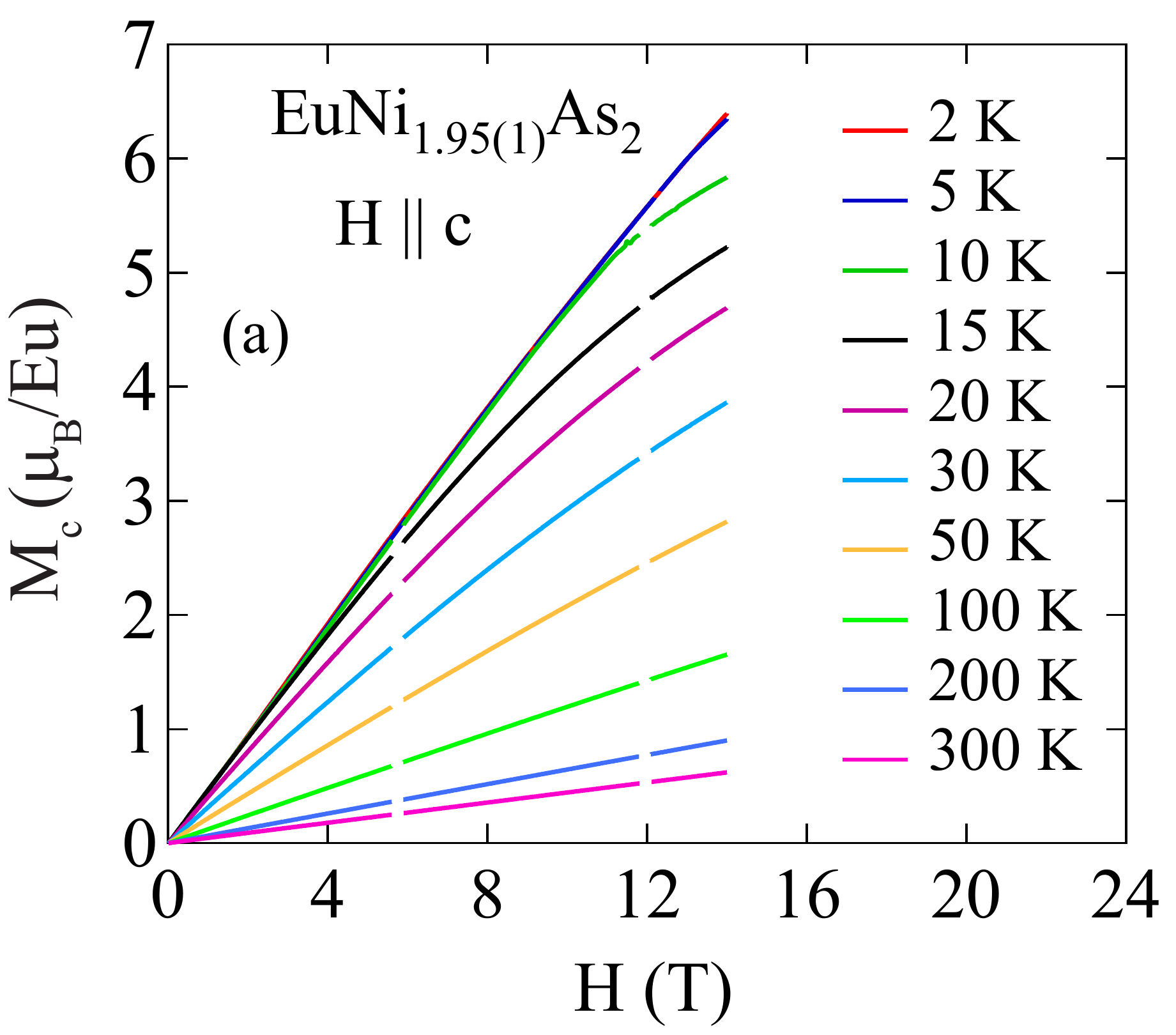}
\includegraphics[width=2.6in]{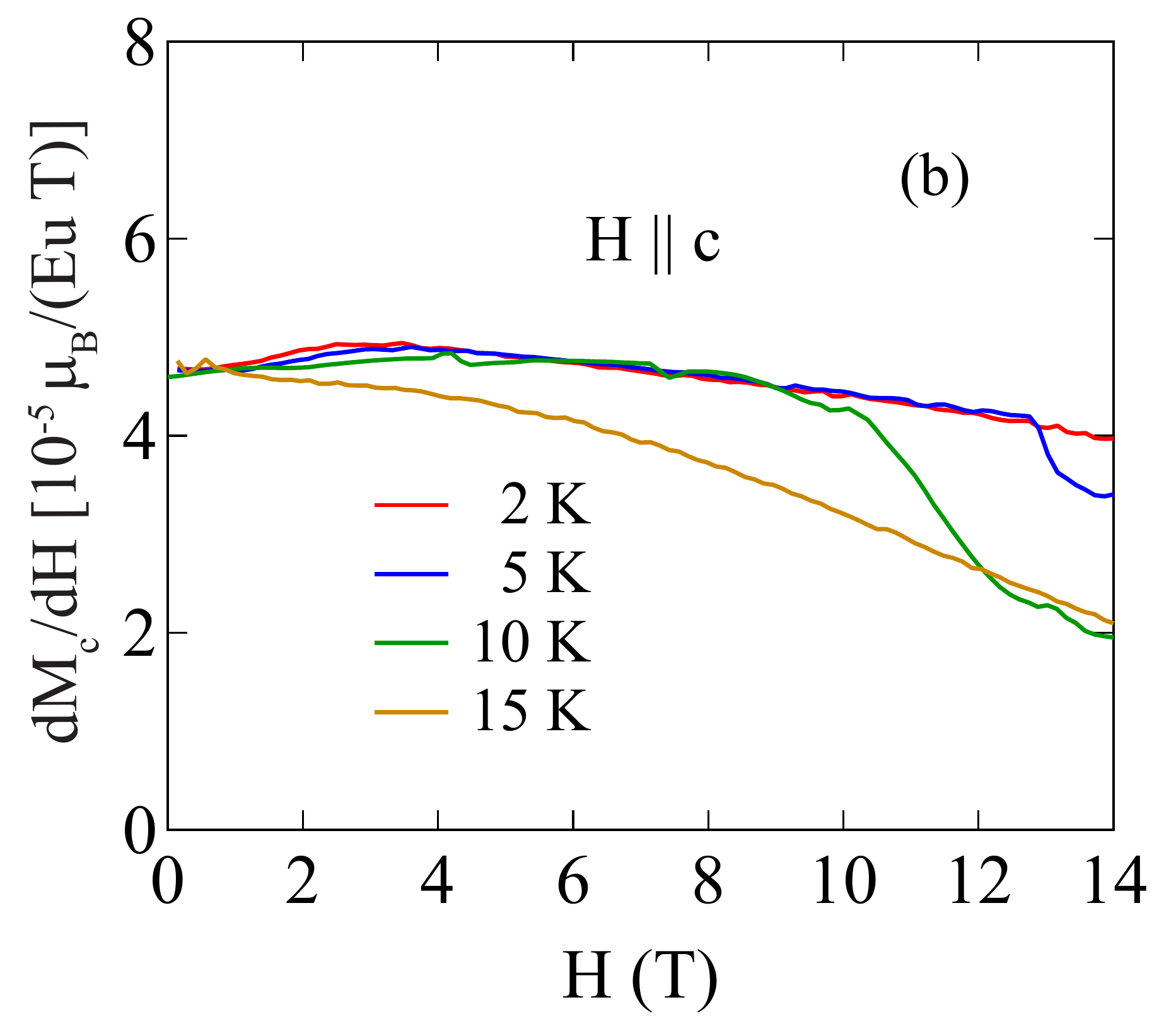}
\caption{Same as Fig.~\ref{Fig:MHab} except with the field applied along the $c$~axis ($H\parallel c$).}
\label{Fig:MHc}
\end{figure}

For a detailed exposition of the high-field behaviors, we obtained $M(H)$ isotherms and their field~derivatives at various temperatures which are plotted in Figs.~\ref{Fig:MHab}(a,b) and \ref{Fig:MHc}(a,b) for $H\parallel ab$ and $H\parallel c$, respectively. The $M_c(H)$ data shown in Fig.~\ref{Fig:MHc}(a) exhibit negative curvature at the higher fields for $T<50$~K. At higher temperatures, a proportional behavior of $M_c(H)$ data is eventually observed. As one can see from the derivative plots in Fig.~\ref{Fig:MHc}(b), the critical field $H_{\rm c\perp}$ at which a second-order AFM to PM transition occurs~\cite{Johnston2015} is shifted to lower field as $T$ increases. The symbol~$\perp$ in $H_{\rm c\perp}$ refers to the critical field with {\bf H} applied perpendicular to the zero-field plane of the ordered moments which is the $ab$~plane in this case.  On the other hand, the $M_{ab}(H)$ isotherms in Fig.~\ref{Fig:MHab}(a) show two clear metamagnetic transitions at $H_{\rm mm1}$ and $H_{\rm mm2}$, respectively, for $T\leq 7$~K that disappear at $T_{\rm N}$. These transitions shift to higher field whereas the critical field $H_{\rm c\perp}$ at the AFM to PM transition shifts to lower field with increasing temperature as seen in the derivative plots in Fig.~\ref{Fig:MHab}(b). The metamagnetic transition fields $H_{\rm mm1}$ and $H_{\rm mm2}$ and the critical field $H_{\rm c\perp}$ are obtained from the fields in the derivative plots at which peaks or discontinuities are seen. The results are listed in Table~\ref{Tab:MH}.

\begin{table}
\caption{\label{Tab:MH} Metamagnetic fields $H_{\rm mm1}$ and $H_{\rm mm2}$, and the critical fields $H_{\rm c\perp}$ with the field perpendicular to the zero-field plane of the ordered moments of EuNi$_{1.95}$As$_2$ single crystals and $H_{\rm c\parallel}$ with the field parallel to the zero-field plane at several temperatures, determined from the isothermal magnetization $M_\alpha(H)$ data in Figs.~\ref{Fig:MHab} and~\ref{Fig:MHc}.}
\begin{ruledtabular}
\begin{tabular}{c|ccc|c}
				&\multicolumn{3}{c|}{$H\parallel ab$} & $H\parallel c$  \\
$T$(K)				&  $H_{\rm mm1}$  & $H_{\rm mm2}$ &  $H_{\rm c\parallel}$	  & $H_{c\perp}$ (T)\\
  	        	 	&    (T)           	 & (T)     		 	& (T)  			&  \\
\hline
2 	  			& 3.15(1) 		& 9.51(3)			& $>14$			&  \\
5	 			 &3.33(2)		&7.73(4)			& $>14$			&12.7(3)\\
7				 & 3.91(2)	 	&5.5(1)			& 13.1(3)			&\\
10				 &			 &				&				& 10.1(6)\\
\end{tabular}
\end{ruledtabular}
\end{table}

Theoretical studies of classical field-induced $xy$-plane ($ab$~plane here) metamagnetic transitions in $z$-axis helices ($c$~axis here) at $T=0$ were presented by one of us for moments confined to the $xy$ plane (infinite $XY$~anisotropy) in Ref.~\cite{Johnston2017b} and more recently for finite XY anisotropy in Ref.~\cite{Johnston2019} where the moments can flop from the $xy$~plane into a three-dimensional spin-flop arrangement on one and/or two spherical ellipses.  Continuous, second-order, and first-order \mbox{metamagnetic} transitions were found, depending on the turn angle~$kd$ and the XY~anisotropy.  For ${\rm EuCo_2P_2}$ with $T_{\rm N} = 66.5$~K and $kd = 0.85\pi$~rad at low temperatures, good fits to the smooth crossover transition between helix and fan phases were obtained by the MFT prediction for both $kd = 6\pi/7~{\rm rad} = 0.857\pi$~rad in Ref.~\cite{Johnston2017b} and $kd = 5\pi/6~{\rm rad} = 0.833\pi$~rad in Ref.~\cite{Johnston2019} with the Eu spins confined to the $ab$~plane by a sufficiently large $XY$ anisotropy.

Here, the low-temperature turn angle for ${\rm EuNi_{1.95}As_2}$ is $kd \approx 0.83\pi$~rad (see following section), which is similar to $kd=0.85\pi$~rad in ${\rm EuCo_2P_2}$.  However, for $M_{ab}(H)$ in Fig.~\ref{Fig:MH2K} for ${\rm EuNi_{1.95}As_2}$ one sees a first-order metamagnetic transition followed by a second-order metamagnetic transition.  In Ref.~\cite{Johnston2019}, we found that if the XY anisotropy is sufficiently smaller than in ${\rm EuCo_2P_2}$, then the sequence of phase transitions observed in ${\rm EuNi_{1.95}As_2}$ can indeed occur when ${\rm EuNi_{1.95}As_2}$ is in the spin-flop phase, as shown in Fig.~4 of the Supplemental Material for Ref.~\cite{Johnston2019} for both $kd = 0.818\pi$ and~$0.833\pi$~rad.  In these cases, a first-order metamagnetic transition arises in the spin-flop phase from a transition from a 3D two-spherical-ellipse ordered-moment configuration to a 3D single-spherical-ellipse configuration, whereas the higher-field second-order metamagnetic transition results from a transition from the 3D single-spherical-ellipse configuration to a single 2D fan configuration in the $ab$~plane.  Finally, at sufficiently high field the system exhibits a second-order transition from the 2D AFM fan phase to the 1D collinear PM phase at $H_{\rm c\parallel}$, where $H_{\rm c\parallel}$ refers to the critical field when {\bf H} is in the plane of the zero-field ordered moments.  From Fig.~\ref{Fig:MHab}(b), one sees that $H_{\rm c\parallel}\approx 13.1$~T at $T=7$~K, as noted in Table~\ref{Tab:MH}.  Thus the occurrence and nature of the two metamagnetic transitions observed in Fig.~\ref{Fig:MH2K} for $H\parallel ab$ can be understood in terms of our model in Ref.~\cite{Johnston2019}.

\subsection{Molecular-Field-Theory Analysis of Magnetic Susceptibility of {\bf EuNi$_{1.95}$As$_2$}}

The anisotropy of $\chi_{ab}(T\leq T_{\rm N})$ with respect to $\chi_c(T\leq T_{\rm N})$ of EuNi$_{1.95}$As$_2$ in Fig.~\ref{Fig:chi} with \mbox{$\chi_{ab}(T\to0)/\chi(T_{\rm N}) \sim 0.6$} suggests intrinsic noncollinear AFM ordering with the ordered moments aligned in the $ab$~plane.  Indeed, the neutron-diffraction study indicated a helical AFM structure with the helix axis being the $c$ axis~\cite{Jin2019} with the ordered moments aligned in the $ab$~plane.  Within MFT, the normalized transverse in-plane susceptibility $\chi_{ab}(T\leq T_{\rm N})/\chi(T_{\rm N})$ of a helical AFM structure is given by~\cite{Johnston2012, Johnston2015}
\newpage
\begin{subequations}
\label{Eqs:Chixy}
\begin{equation}
\frac{\chi_{ab}(T \leq T_{\rm N})}{\chi(T_{\rm N})}=  \frac{(1+\tau^*+2f_J+4B^*)(1-f_J)/2}{(\tau^*+B^*)(1+B^*)-(f_J+B^*)^2},
\label{eq:Chi_plane}
\end{equation}
 where
\begin{equation}
B^*=  2(1-f_J)\cos(kd)\,[1+\cos(kd)] - f_J,
\label{eq:Bstar}
\end{equation}
\begin{equation}
t =\frac{T}{T_{\rm N}},\quad \tau^*(t) = \frac{(S+1)t}{3B'_S(y_0)}, \quad y_0 = \frac{3\bar{\mu}_0}{(S+1)t}, 
\end{equation}
$f_J = f= \theta_{\rm p}/T_{\rm N}$ in Table~\ref{Tab.chidata} for weak anisotropy as in ${\rm EuNi_{1.95}As_2}$, the ordered moment versus $T$ in $H=0$ is denoted by $\mu_0$, and the reduced ordered moment $\bar{\mu}_0 = \mu_0/\mu_{\rm sat}$ is determined by numerically solving the self-consistency equation
\begin{equation}
\bar{\mu}_0 = B_S(y_0).
\label{Eq:barmuSoln}
\end{equation}
\end{subequations}
We define
\begin{equation}
B'_S(y_0) \equiv \frac{dB_S(y)}{dy}\bigg|_{y=y_0},
\end{equation}
where the Brillouin function $B_S(y)$ is 
\begin{equation}
B_S(y)= \frac{1}{2 S}\left\{(2S+1){\rm coth}\left[(2S+1)\frac{y}{2}\right]-{\rm coth}\left(\frac{y}{2}\right)\right\}.
\label{Eq:BSy}
\end{equation}
Here $kd = k(2\pi/c)(c/2)= k\pi$~rad is the turn angle between the ordered magnetic moments in adjacent layers along the helix axis, where the AFM propagation vector is written as $(0,0,k)2\pi/c$. At $T=0$, Eqs.~(\ref{Eqs:Chixy}) yield~\cite{Johnston2012,Johnston2015} 
\begin{equation}
\label{Eq:kd}
\frac{\chi_{ab}(T=0)}{\chi(T_{\rm N)}}=\frac{1}{2[1+2~{\rm cos}(kd)+2~{\rm cos}^2(kd)]}.
\end{equation}

\begin{figure}[h]
\includegraphics[width=2.75in]{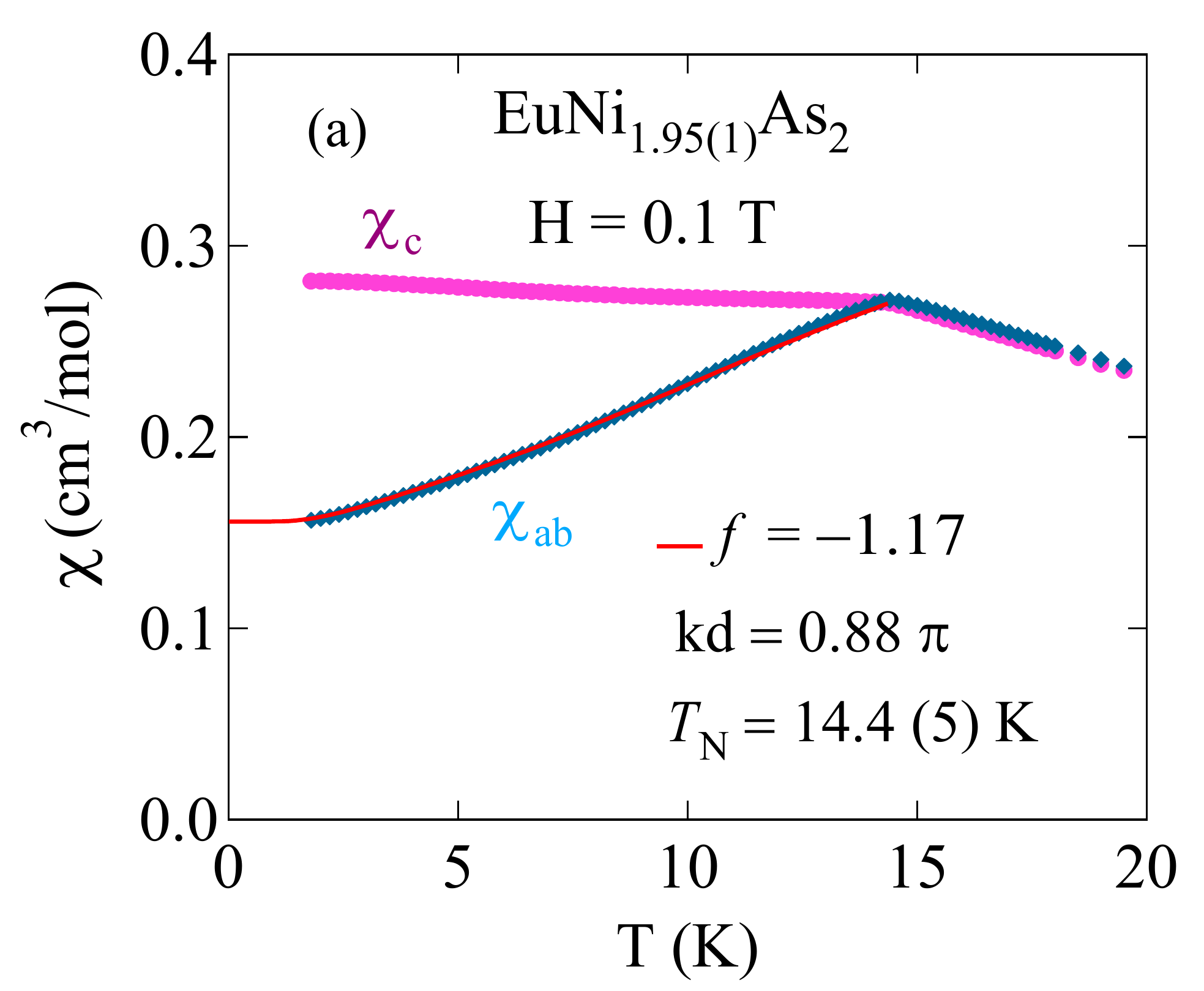}
\includegraphics[width=2.75in]{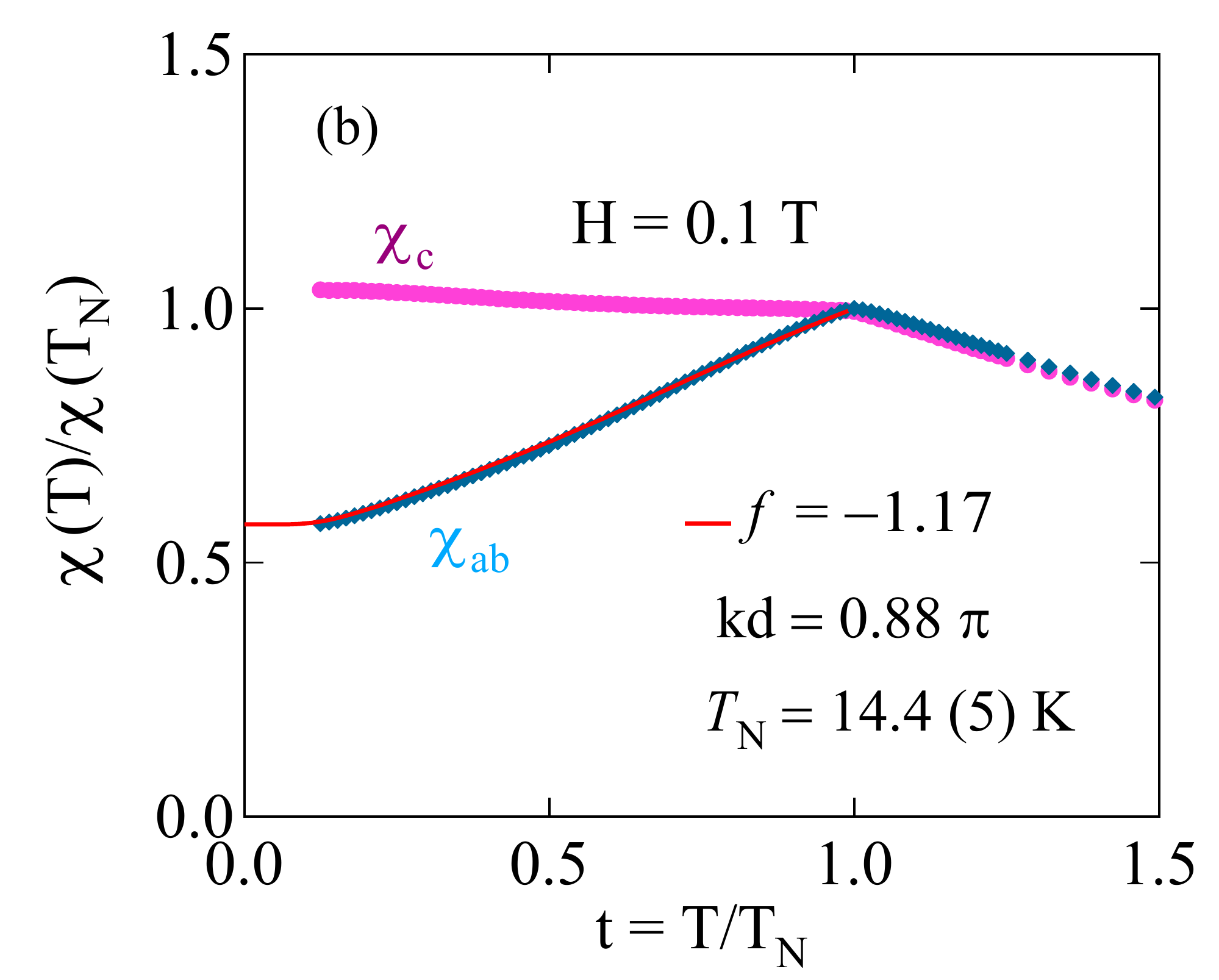}
\caption{(a)~Magnetic susceptibility $\chi$ versus temperature~$T$ for fields of magnitude $H=0.1$~T parallel ($\chi_c$) and perpendicular ($\chi_{ab}$) to the tetragonal $c$~axis of single-crystal EuNi$_{1.95}$As$_2$ at low temperatures.  (b)~The $\chi(T)$ data in~(a) normalized by $\chi(T_{\rm N})$.  The fit of $\chi_{ab}(T)/\chi(T_{\rm N})$ in~(b) and of $\chi_{ab}(T)$ in~(a) for $T\leq T_{\rm N}$ by the MFT prediction in Eqs.~(\ref{Eqs:Chixy}) for a helix is shown as the solid curves. }
\label{Fig:ChiTNN}
\end{figure}

The $\chi(T)$ data in Fig.~\ref{Fig:chi} are plotted on an expanded temperature scale in Fig.~\ref{Fig:ChiTNN}(a). The scaled data $\chi(T)/\chi(T_{\rm N})$ necessary for fitting the $\chi_{ab}(T)$ data by Eqs.~(\ref{Eqs:Chixy}) are plotted in Fig.~\ref{Fig:ChiTNN}(b).  For the fits we used $S=7/2$ and the value $f=-1.17$ from the average of the two values for $f_{ab}$ with $H=0.1$~T in Table~\ref{Tab.chidata}. Taking $\chi_{ab}(T=0)/\chi(T_{\rm N}) = 0.57$ from Fig.~\ref{Fig:ChiTNN}(b) and solving Eq.~(\ref{Eq:kd}) for $kd$ gives the two solutions $kd=0.52\pi$~rad $(94^\circ)$ and $kd=0.88\pi$~rad $(158^\circ)$ for the helix turn angle for EuNi$_{1.95}$As$_2$ at $T=0$. In order to fit the \mbox{lowest-$T$} $\chi_{ab}$ data in Fig.~\ref{Fig:ChiTNN}(b), we used $kd(T=0)=0.88\pi$~rad which is in reasonable agreement with the value from the neutron diffraction measurement ($kd=0.9200\pi$~rad)~\cite{Jin2019}. The \mbox{$\chi_{ab}(T\leq T_{\rm_N})/\chi(T_{\rm N})$} fit thus obtained using Eqs.~(\ref{Eqs:Chixy}) is plotted as the solid red curve in Fig.~\ref{Fig:ChiTNN}(b).  The corresponding fit on the absolute susceptibility scale is shown in Fig.~\ref{Fig:ChiTNN}(a). The fit is seen to be in excellent agreement with the experimental $\chi_{ab}(T)$ data.

\begin{table*}
\caption{\label{Tab:HEI} Exchange constants $J_0$, $J_1$, and $J_2$ in Fig.~\ref{Fig:J0_J1_J2_model_helix} and their sum $J_{\rm tot}$ obtained by solving Eqs.~(\ref{Eqs:J0J1zJ2z}).  Also listed are the exchange constants between Eu spins $J{\rm_A}$, $J{\rm_B}$ and $J{\rm_C}$ in Fig.~\ref{Fig:JA_JB_JC_model} obtained from Eqs.~(\ref{eq:Jabc}). Negative $J$ values are FM and positive values are AFM. Also shown are the Weiss temperature $\theta_{\rm p}$ in the Curie-Weiss law calculated from Eq.~(\ref{Eq.chi}) and the listed $J$ values. The sum of the $J$ values is $J_{\rm tot} = J_0 + 2 J_{1}+ 2 J_{2}$ and the $\theta_{\rm p}$ values are obtained using Eq.~(\ref{eq:thetap}). The references are listed in the last column, where PW means the present work.}
\begin{ruledtabular}
\begin{tabular}{llcccccccccc}
Compound
	& $kd$	
	& $J_0$ 
	& $J_1$ 
	& $J_2$ 
	& $J_{\rm tot}$ 
	& $J{\rm_A}/k{\rm_B}$
	& $J{\rm_B}/k{\rm_B}$
	& $J{\rm_C}/k{\rm_B}$  
	& $J{\rm_B}/J{\rm_C}$ 
	&$\theta{\rm_p}$
	& Ref.\\
        & (rad) &(K)     &(K)         &(K)               &(K)        &(K)         &(K)               &(K)                        &(K) & (K)\\
\hline
${\rm EuCo_2P_2}$ 	& $0.857\pi$	& $-9.55$		&2.14	&0.594	& $-4.08$				& $-2.39$	 	& 0.535	& 0.594	& 0.90 	& 21.5  	&\cite{Sangeetha2016}\\
${\rm EuCo_2As_2}$ 	& $0.79\pi$	& $-6.85$		&1.22	&0.387	& $-3.63$				& $-1.712$	& 0.306	& 0.387	& 0.7901 	& 19.07 	&\cite{Sangeetha2018}\\
EuNi$_2$As$_2$ 	& $0.88\pi$	& $-0.86$		&1.29	&0.348	& 2.42				& $-0.216$	& 0.323	& 0.348	& 0.929 	& $-12.7$ & PW\\
${\rm EuRh_2As_2}$ 	& $0.83\pi$	& $-6.20$		&1.91	&0.554	& $-1.27$				& $-1.55$	 	& 0.477	& 0.554	& 0.86 	& 6.76  &PW\\
\end{tabular}
\end{ruledtabular}
\end{table*}

\subsection{\label{Sec:HeisExchInts} Heisenberg Exchange Interactions in ${\rm\bf EuCo_2As_2}$}

\begin{figure}
\includegraphics [width=2in]{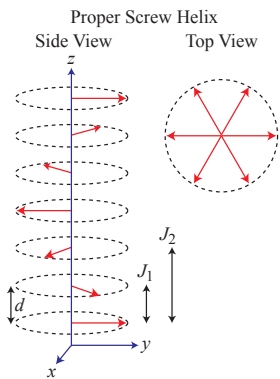}
\caption {Generic helical AFM structure \cite{Johnston2012}.  Each arrow represents a layer of moments perpendicular to the $z$~axis that are ferromagnetically aligned within the $xy$ plane and with interlayer separation $d$.  The wave vector {\bf k} of the helix is directed along the $z$~axis.  The magnetic moment turn angle between adjacent magnetic layers is $kd$, where $k$ is the magnitude of the wave vector.  The top view shows the magnetic moments as viewed from the positive $z$~axis.  When the moment vectors are placed tail to tail as shown, the picture is a hodograph of the magnetic moments.  The MFT nearest-layer and next-nearest-layer exchange interactions $J_1$ and $J_2$ are indicated.}
\label{Fig:J0_J1_J2_model_helix}
\end{figure}

We now estimate the Eu intralayer and the Eu iinterlayer Heisenberg exchange interactions along the $c$~axis within the minimal $J_0$-$J_{1}$-$J_{2}$ MFT model for a helix \cite{Johnston2015, Johnston2019, Nagamiya1967}, where $J_0$ is the sum of the Heisenberg exchange interactions of a representative spin with all other spins in the same spin layer perpendicular to the helix ($z=c$) axis, $J_{1}$ is the sum of all interactions of the spin with spins in a nearest layer along the helix axis, and $J_{2}$ is the sum of all interactions of the spin with spins in a second-nearest layer, as indicated in Fig.~\ref{Fig:J0_J1_J2_model_helix}.  Within this one-dimensional model, $kd$, $T_{\rm N}$, and $\theta_{{\rm p}}$ are related to these exchange interactions by \cite{Johnston2012, Johnston2015}.
\begin{subequations}
\label{Eqs:J0J1zJ2z}
\begin{eqnarray}
&&\cos(kd) = -\frac{J_{1}}{4J_{2}},\label{Eq:coskdJ1J2}\\*
T_{\rm N} &=& -\frac{S(S+1)}{3k_{\rm B}} \big[J_0 + 2J_{1}\cos(kd)\label{eq:TN}\\*
&& \hspace{0.9in} +\ 2J_{2}\cos(2kd)\big], \nonumber\\*
\theta_{\rm p} &=& -\frac{S(S+1)}{3k_{\rm B}} \left(J_0+2J_{1}+2J_{2}\right), \label{eq:thetap}
\end{eqnarray}
\end{subequations}
where a positive (negative) $J$ corresponds to an AFM (FM) interaction.  Using $S = 7/2$,  $T_{\rm N}=14.5$~K, \mbox{$\theta_{\rm p}=\theta_{\rm p\,ave}$} in Table~\ref{Tab.chidata}, and $kd=0.88\pi$~rad, solving Eqs.~(\ref{Eqs:J0J1zJ2z}) for $J_0$, $J_1$, and $J_2$ yields the values listed in Table~\ref{Tab:HEI}.  Table~\ref{Tab.chidata} gives a negative (AFM-like) $\theta_{\rm p}$ value, consistent with the net interaction strength $J_{\rm tot} =J_0+2J_{1}+2J_{2}$ being AFM-like.  Indeed, the interplane exchange constants $J_{1}$ and $J_{2}$ must both be positive (AFM-like) in order to obtain the observed helical AFM structure with $\pi/2<kd\leq \pi$~rad.  The FM (negative) intralayer interaction $J_0$ is required so that the Eu spins in each $ab$-plane layer are ferromagnetically aligned within the helical structure.

\begin{figure}
\includegraphics[width=1in]{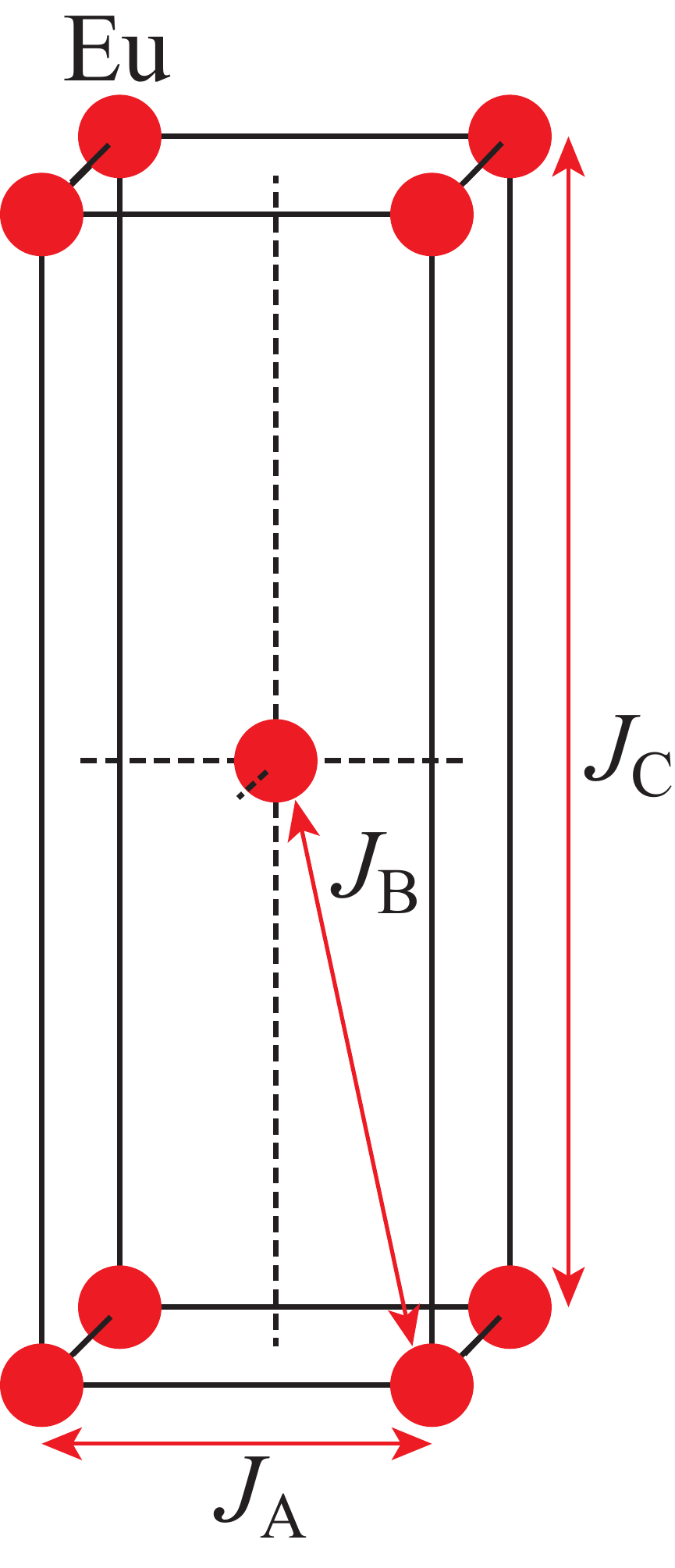}
\caption {Exchange interactions $J_{\rm A}$, $J_{\rm B}$, and $J_{\rm C}$ between the Eu spins in a body-centered-tetragonal unit cell of EuNi$_{1.95}$As$_2$.}
\label{Fig:JA_JB_JC_model}
\end{figure}

Estimates of the Heisenberg exchange interactions $J\rm_A$, $J\rm_B$, and $J\rm_C$ between the Eu spins in the body-centered-tetragonal unit cell of EuNi$_{1.95}$As$_2$ as shown in Fig.~\ref{Fig:JA_JB_JC_model} can be found from the $J_0$, $J_{1}$, and $J_{2}$ values in Table~\ref{Tab:HEI} according to
\begin{equation}
J_0 = 4J_{\rm A}, \quad J_{1} = 4J_{\rm B}, \quad J_{2} = J_{\rm C},
\label{eq:Jabc}
\end{equation}
and the results are listed in Table~\ref{Tab:HEI}. Also shown are the helix turn angles and exchange constants estimated via MFT for other related Eu-based 122-type helical antiferromagnets including ${\rm EuRh_2As_2}$ which we analyze in the following section.

\subsection{Molecular-Field-Theory Analysis of Magnetic Susceptibility of ${\rm\bf EuRh_2As_2}$}

\begin{figure}
\includegraphics[width=2.75in]{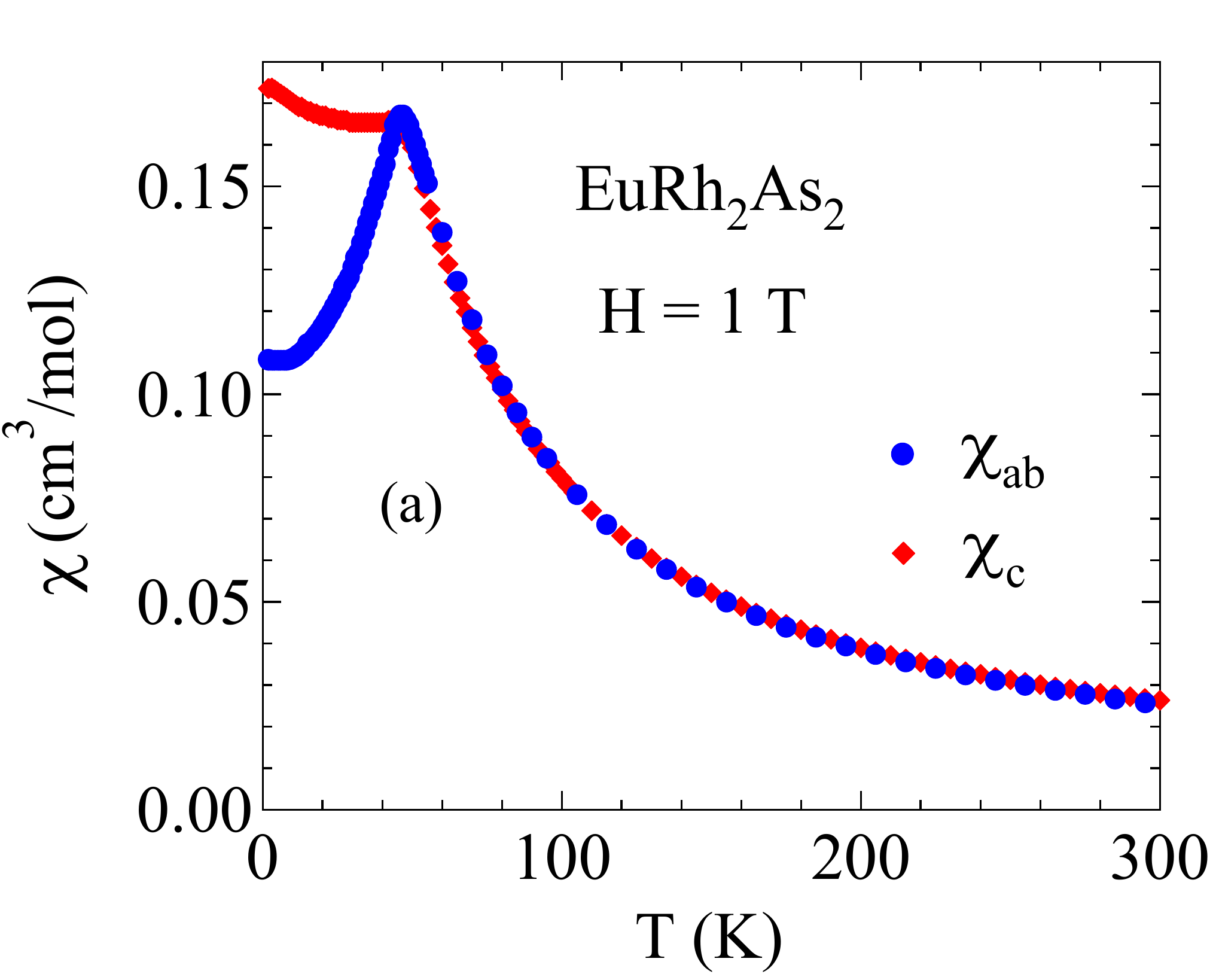}
\includegraphics[width=2.7in]{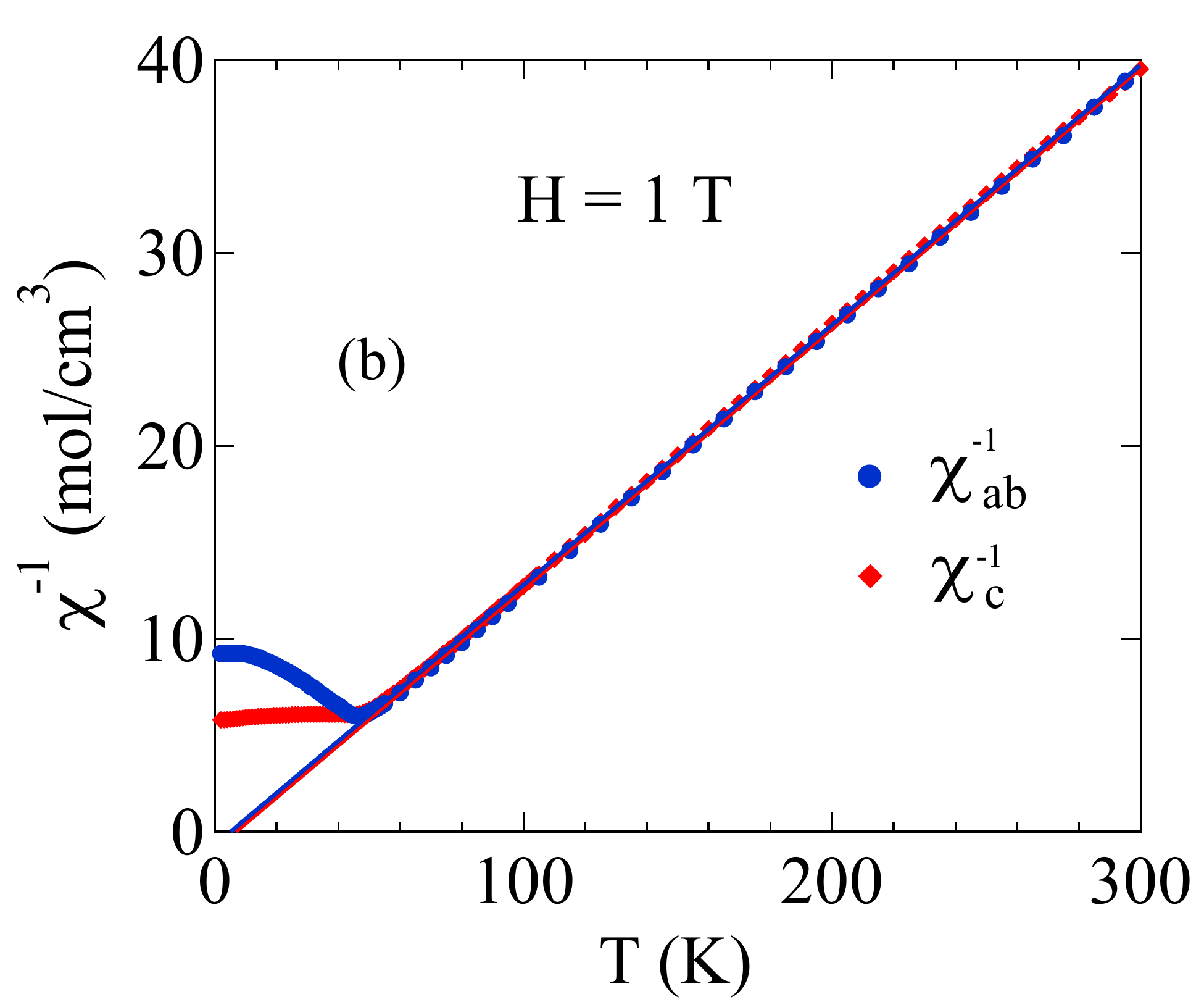}
\caption{(a) $\chi_{ab}$ and $\chi_{c}$ as a function of temperature $T$ in an applied magnetic field $H=1$~T and (b) $\chi^{-1}(T)$ and its Curie-Weiss fit by Eq.~(\ref{Eq.cw}) for the helical antiferromagnet ${\rm EuRh_2As_2}$ obtained using the data in Ref.~\cite{Singh2009}.}
\label{Fig:magERA}
\end{figure}

Anisotropic magnetic susceptibility data for single-crystal ${\rm EuRh_2As_2}$ revealed AFM ordering at $T_{\rm N}=47$~K~\cite{Singh2009}, as shown in Fig.~\ref{Fig:magERA}(a). The Curie-Weiss fit to the inverse susceptibility data $\chi^{-1}(T)$ for both field orientations is shown in Fig.~\ref{Fig:magERA}(b) and the fitted Curie constant and Weiss temperature are $C = 7.42(4)$~cm$^3$\,K/mol and $\theta_{\rm p} = 6.76(5)$~K, respectively.  Eu valence fluctuations have been suggested~\cite{Singh2009} as the reason for the smaller value of the Curie-constant in ${\rm EuRh_2As_2}$ compared to the value 7.88~cm$^3$\,K/mol for $S=7/2$ with $g=2$. The  positive value of $\theta_{\rm p}$ indicates a net FM exchange interaction between the Eu$^{+2}$ magnetic moments.  X-ray resonant magnetic scattering measurements on a single crystal showed that the predominant magnetic structure is a $c$-axis helix with a turn angle of $\approx 0.9\pi\,{\rm rad}\approx 160^\circ$~\cite{Nandi2009}.  Therefore we analyze here the anisotropic magnetic susceptibility literature data for ${\rm EuRh_2As_2}$~\cite{Singh2009} in terms of a helical model within MFT\@.

\begin{figure}
\includegraphics[width=2.7in]{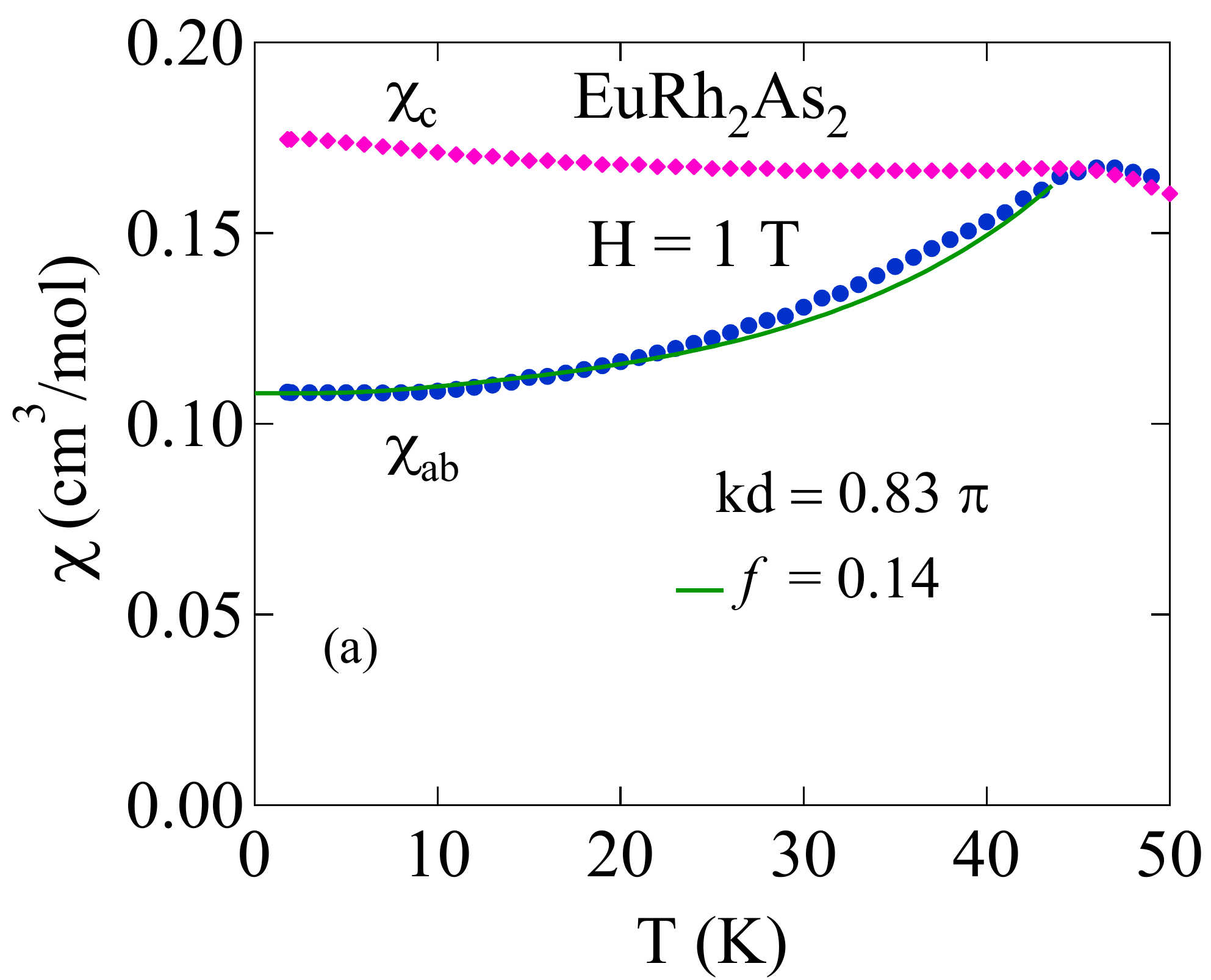}
\includegraphics[width=2.7in]{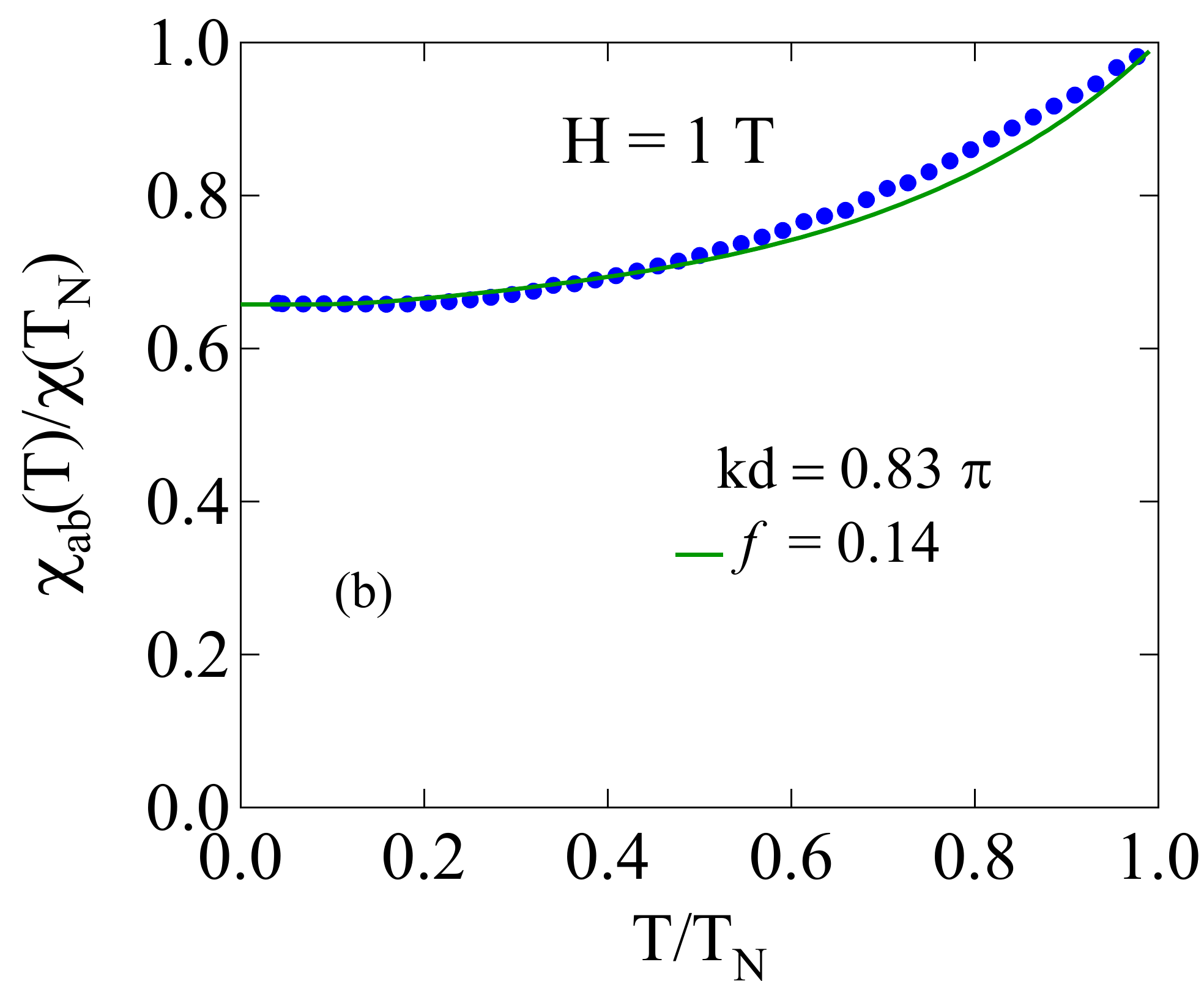}
\caption{(a)~Magnetic susceptibility $\chi$ versus temperature~$T$ for fields of magnitude $H=1$~T parallel ($\chi_c$) and perpendicular ($\chi_{ab}$) to the tetragonal $c$~axis of single-crystal ${\rm EuRh_2As_2}$ at low temperatures~\cite{Singh2009}.  (b)~The $\chi(T)$ data in~(a) normalized by $\chi(T_{\rm N})$.  The fit of $\chi_{ab}(T)/\chi(T_{\rm N})$ in~(b) and of $\chi_{ab}(T)$ in~(a) for $T\leq T_{\rm N}$ by the MFT prediction for a helix with $kd=0.83\pi$ and $f=~0.14$ in Eqs.~(\ref{Eqs:Chixy}) for a helix are shown as the solid curve.}
\label{Fig:ChiTNN_ERA}
\end{figure}
The value of $\chi_{ab}(T\rightarrow0)/\chi(T_{\rm N})$ in Fig.~\ref{Fig:magERA}(a) is consistent with ${\rm EuRh_2As_2}$ being a helical $c$-axis antiferromagnet with the ordered moments oriented within the $ab$ plane. Expanded plots of the data in Fig.~\ref{Fig:magERA} at low temperatures are shown in Fig.~\ref{Fig:ChiTNN_ERA}. Using the value $\chi_{ab}(T=0)/\chi(T_{\rm N}) = 0.66$ from Fig.~\ref{Fig:ChiTNN_ERA}(b) and solving Eq.~(\ref{Eq:kd}) for $kd$ gives the two solutions $kd=0.54\pi$\,rad $(98^\circ)$ and $kd=0.83\pi$\,rad $(149^\circ)$ for the helix turn angle at $T=0$. We used $kd(T=0)=0.83\pi$\,rad for ${\rm EuRh_2As_2}$ because this $kd$  is comparable to the value of $kd$ from Ref.~\cite{Nandi2009} as well as neutron diffrraction studies on similar compounds as listed in Table~\ref{Tab:HEI}. The scaled data $\chi_{ab}(T\leq T_{\rm_N})/\chi(T_{\rm N})$ are fitted by Eqs.~(\ref{Eqs:Chixy}) using $S=7/2$, $T_{\rm N}=47$~K, $f=6.5/47=0.14$ and $kd=0.83\pi$\,rad as shown by the solid curve in Fig.~\ref{Fig:ChiTNN_ERA}(b). The corresponding fit is shown in Fig.~\ref{Fig:ChiTNN_ERA}(a) as another solid curve. The fits are seen to be quite good.  The Heisenberg exchange interactions between the Eu spins in ${\rm EuRh_2As_2}$ were derived from Eqs.~(\ref{Eqs:J0J1zJ2z}) and~(\ref{eq:Jabc}) using $S=7/2$ and $kd=0.83\pi$\,rad and are listed in Table~\ref{Tab:HEI}.

\section{\label{Sec:Res} Electrical Resistivity}

\begin{figure}
\includegraphics[width=2.75in]{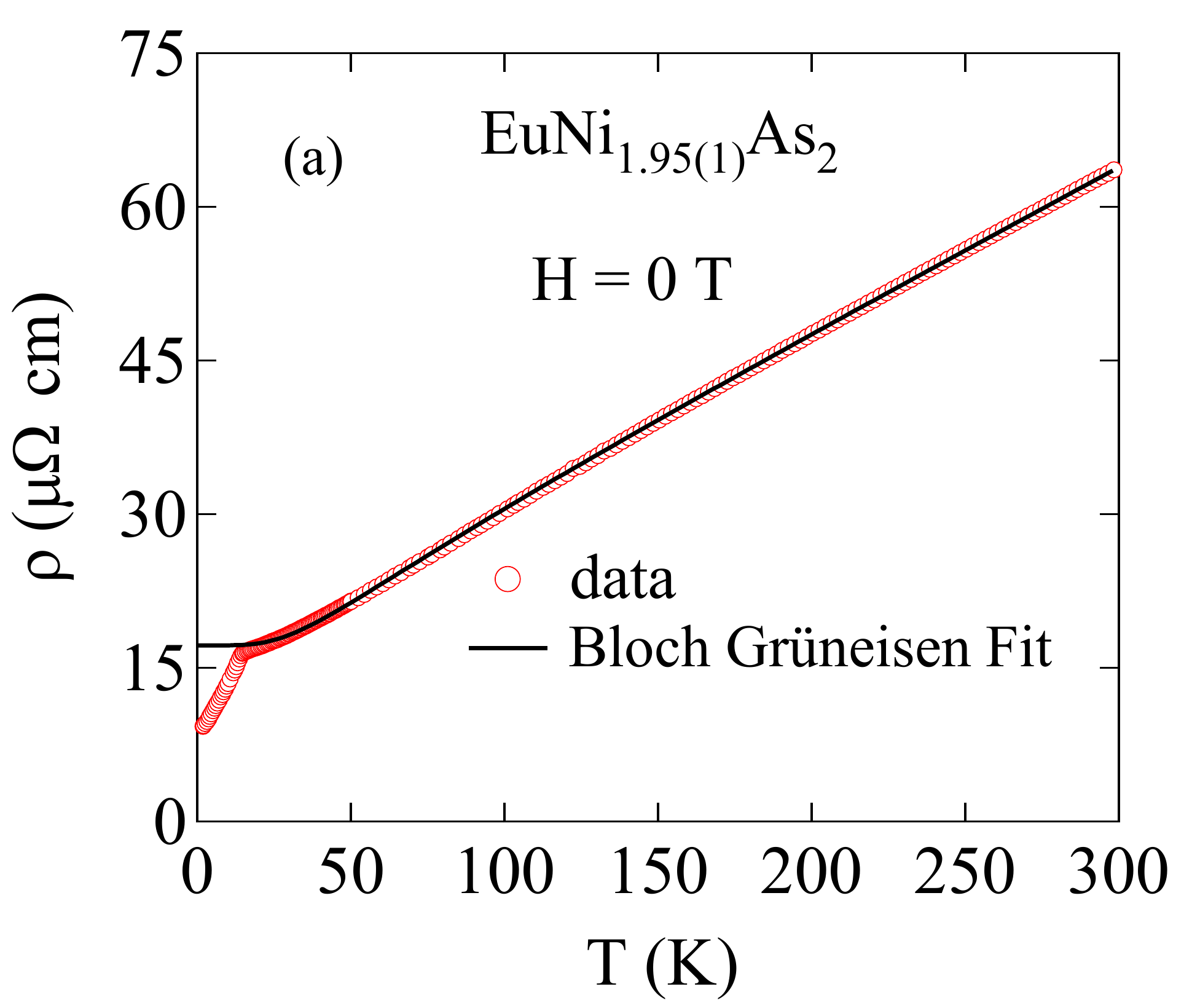}
\includegraphics[width=2.75in]{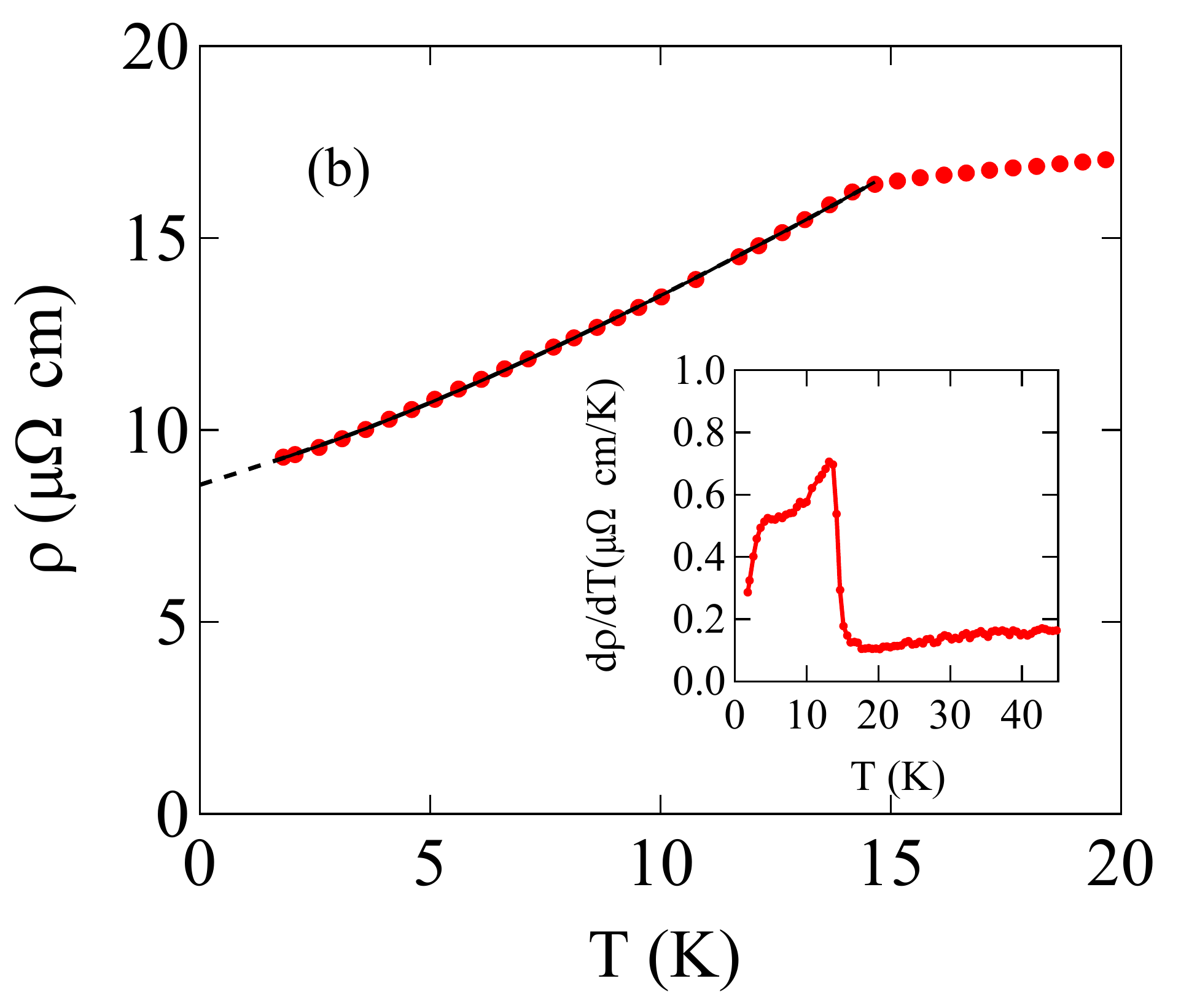}
\caption{(a)~In-plane electrical resistivity $\rho$ of  ${\rm EuNi_{1.95}As_2}$ as a function of temperature from 1.8 to 300~K measured in zero magnetic field. The solid black curve is the fit of the prediction of the Bloch-Gr\"uneisen theory in Eqs.~(\ref{Eqs:rhoBG}) to the data above 20~K and is extrapolated to $T=0$. (b)~Expanded plots of $\rho(T)$ at low temperature. The black line is the fit of the data by Eq.~(\ref{Eq:rhoFitTn}) over the temperature interval $2~{\rm K}\leq T\leq 14$~K with a dashed-line extrapolation to~0~K\@. Inset: temperature derivative $d\rho/dT$ versus~$T$\@.}
\label{Fig:Res0T}
\end{figure}

The in-plane ($ab$-plane) electrical resistivity $\rho$ of an EuNi$_{1.95}$As$_2$ crystal  as a function of temperature $T$ from 1.8 to 300~K measured at zero magnetic field is shown in Fig.~\ref{Fig:Res0T}(a). The $\rho(T)$ exhibits metallic behavior. The residual resistivity at $T=1.8$~K is $\rho_0\approx 9.3~\mu\Omega$~cm and the residual resistivity ratio is RRR~$\equiv \rho(300~{\rm K})/\rho(1.8~{\rm K})\approx6.9$.  The AFM transition is observed at  $T_{\rm N} = 14.2(8)$~K, as clearly shown in a plot of the derivative $d\rho$($T$)/$dT$ versus $T$ in Fig.~\ref{Fig:Res0T}(b) inset, consistent with the $T_{\rm N}$ values found from our $\chi(T)$ data above and $C_{\rm p}(T)$ data below. In addition, the data in the inset show a slope change at $\approx4.0(5)$~K of unknown origin.

The low-$T$ $\rho_{ab}(T)$ data below $T_{\rm N}$ are fitted from 2~K to 14~K by
\begin{equation}
\rho(T)=\rho_0+AT^n,
\label{Eq:rhoFitTn}
\end{equation}
as shown by the solid curve in Fig.~\ref{Fig:Res0T}(b). The fitted parameters are $\rho_0=8.71(3)~\mu \Omega{\rm~cm}$, $A=0.262(9)~\mu \Omega{\rm~cm/K}^n$ with $n=1.26(1)$.  Thus $\rho(T\leq T_{\rm N})$ does not follow Fermi-liquid $T^2$ behavior, likely because it is affected by the $T$-dependent loss of spin-disorder scattering on cooling below $T_{\rm N}$.

The $\rho(T)$ in the normal state above 20~K is fitted by the Bloch-Gr\"{u}neisen (BG) model where the resistivity arises from scattering of electrons from acoustic phonons.  Our fitting function is 
\begin{subequations}
\label{Eqs:rhoBG}
\begin{equation}
\rho{\rm_{BG}}(T)=\rho_0+\rho_{\rm sd}(T) +C~f(T/\Theta_R),
\end{equation}
where~\cite{Goetsch2012}
\begin{equation}
f(T/\Theta_R) = \left(\frac{T}{\Theta_R}\right)^5\int_{0}^{\Theta_{\rm D}/T}\frac{x^5 dx}{(1-e^{-x})(e^x-1)}dx.
\label{Eq:BGM}
\end{equation}
\end{subequations}
Here $\rho_0+\rho_{\rm sd}(T)$ is the sum of the residual resistivity $\rho_0$  due to static defects in the crystal lattice and the spin-disorder resistivity $\rho_{\rm sd}(T)$.  The constant $C$ describes the $T$-independent interaction strength of the conduction electrons with the thermally excited phonons and contains the ionic mass, Fermi velocity, and other parameters, $x=\frac{h\omega}{2\pi k_BT}$, and $\Theta_R$ is the Debye temperature determined from electrical resistivity data. The representation of $f(T/\Theta_R)$ used here is an accurate analytic Pad\'{e} approximant function of $T/\Theta_R$ determined by a fit to the integral on the right-hand side of Eq.~(\ref{Eq:BGM}) \cite{Goetsch2012}. The fit to the $\rho(T)$ data between 20 and 300~K by Eqs.~(\ref{Eqs:rhoBG}) is shown as the solid black curve in Fig.~\ref{Fig:Res0T}(a). The fitted parameters are $(\rho_0+\rho_{\rm sd})=17.18(3)~\mu\Omega\,{\rm cm}$, $C=27.0(2)~\mu\Omega\,{\rm cm}$ and $\Theta_R = 179(1)$~K\@. The $\rho_{\rm sd}$ calculated from the value of  $(\rho_0+\rho_{\rm sd})$ using $\rho_0(1.8~\rm K) = 9.29~\mu \Omega$\,cm is \mbox{$\rho_{\rm sd}\approx 7.89~\mu \Omega\,{\rm cm}$}.

\begin{figure}
\includegraphics[width=2.75in]{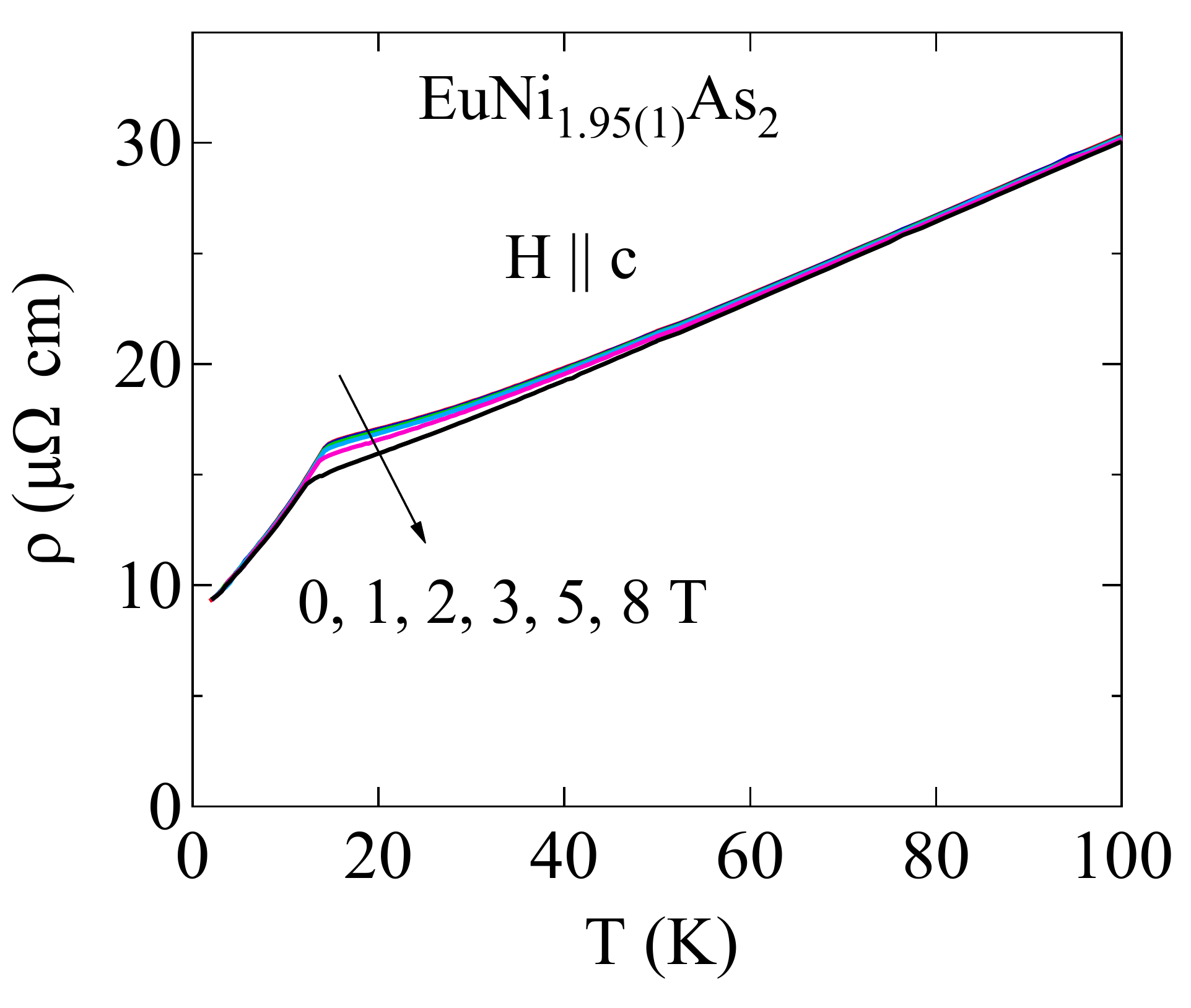}
\caption{In-plane electrical resistivity $\rho$ of ${\rm EuNi_{1.95}As_2}$ function of temperature between 1.8 to 300~K measured in the indicated magnetic fields $H\parallel c$.}
\label{Fig:ResH}
\end{figure}

The $\rho(T)$ of single-crystal EuNi$_{1.95}$As$_2$ in various magnetic fields applied along the $c$~axis is shown in Fig.~\ref{Fig:ResH}. We find that $T_{\rm N}$ decreases from 14.2~K at $H=0$ to 12.3~K at $H=8$~T, as plotted in Fig.~\ref{Fig:PD} below.  The $\rho(T>T_{\rm N})$ shows negative magnetoresistance, with the strongest dependence just above $T_{\rm N}$.

\section{\label{Sec:HC}  Heat capacity}

\begin{figure}
\includegraphics[width=2.5in]{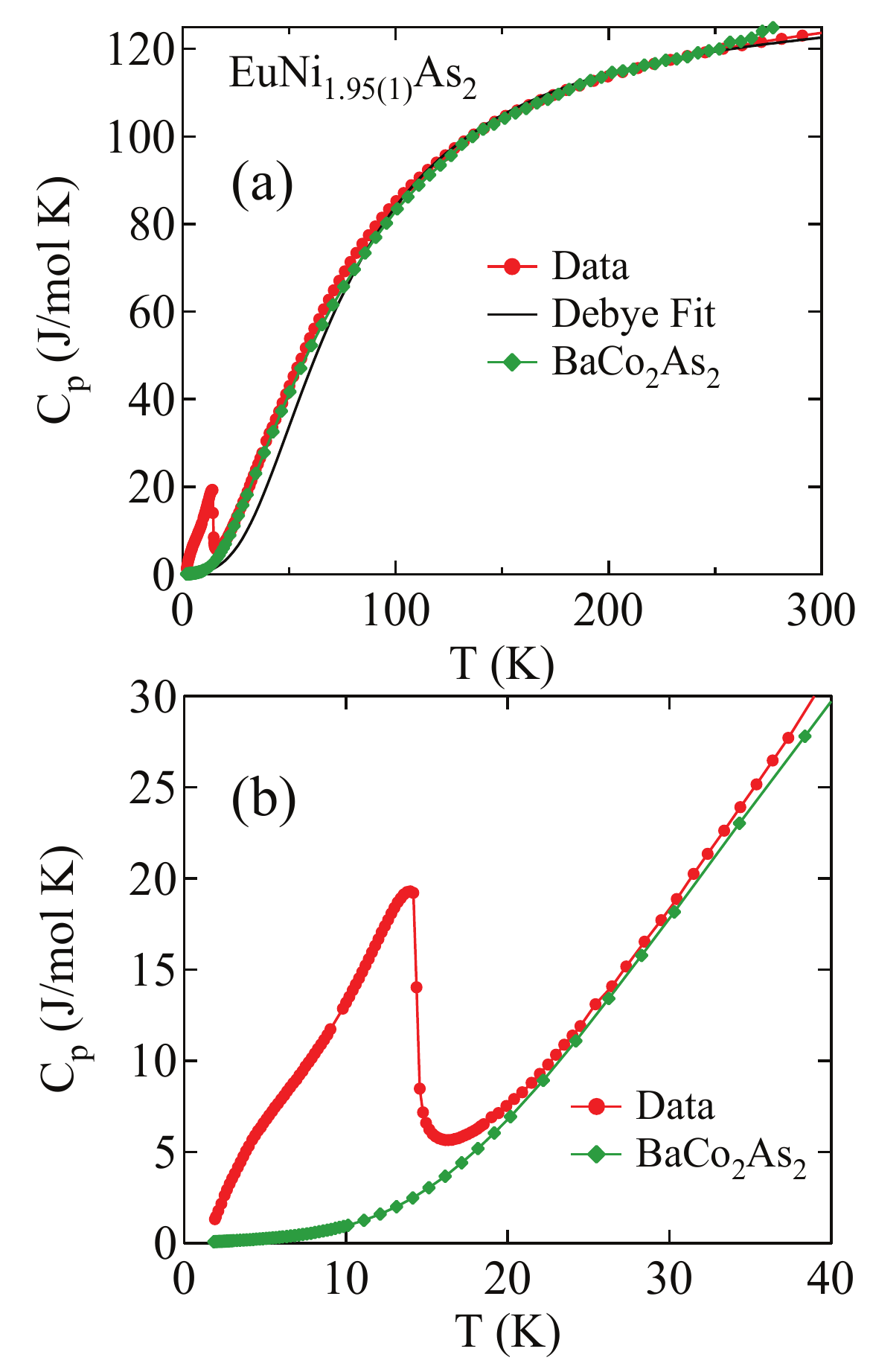}
\caption{(a) Temperature~$T$ dependence of the heat capacity $C_{\rm p}$ for ${\rm EuNi_{1.95}As_2}$ and ${\rm BaCo_2As_2}$~\cite{Sangeetha2018} single crystals in $H = 0$~T\@. The black solid curve is a fit of the data between 50 and 300~K by the Debye lattice heat capacity model in Eqs.~(\ref{Eq:Cpdebye}). (b) Expanded plot of $C_{\rm p}(T)$ for EuNi$_{1.95}$As$_2$ and ${\rm BaCo_2As_2}$ at low temperature.}
\label{Fig:cp0T}
\end{figure}

\begin{figure}
\includegraphics[width=2.5in]{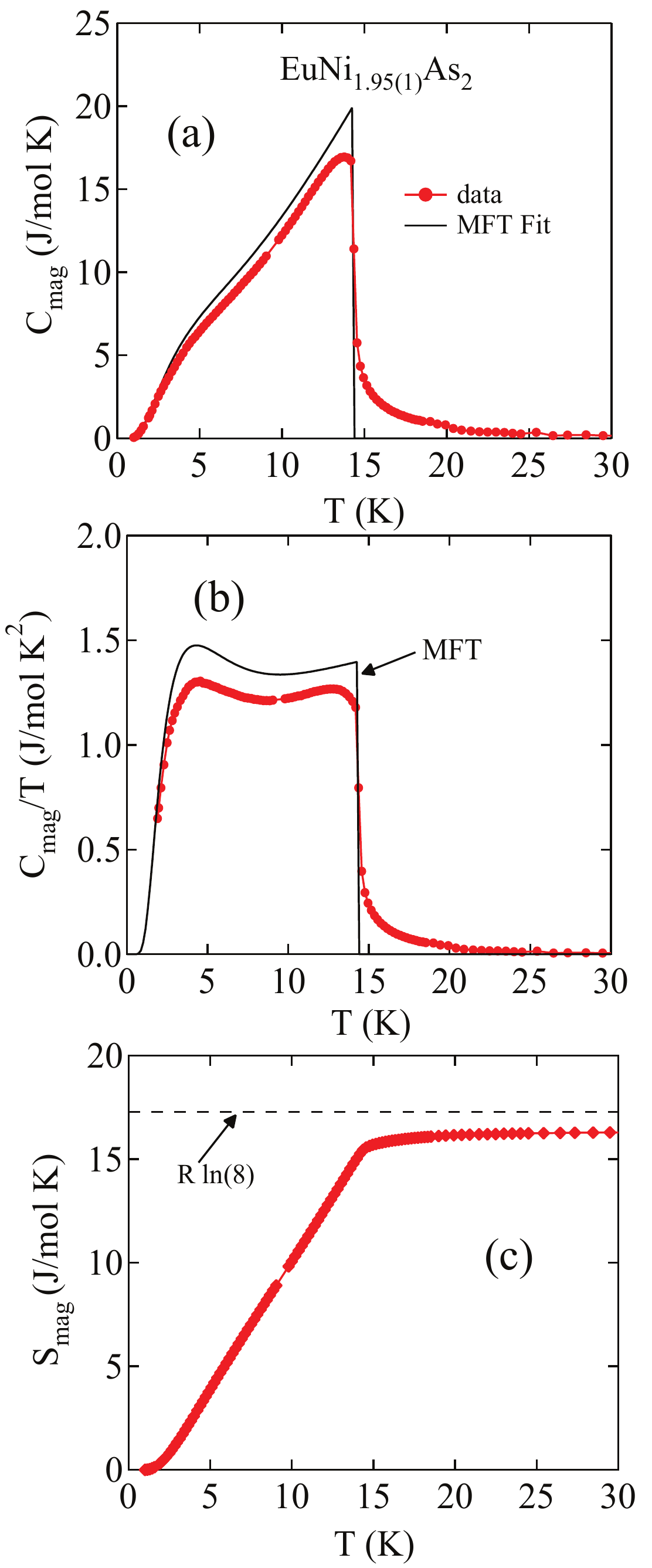}
\caption{(a) Magnetic contribution $C_{\rm mag}$ of EuNi$_{1.95}$As$_2$ versus temperature~$T$ between 1.8 and 30~K obtained by subtracting the nonmagnetic contribution $C_{\rm p}(T)$ of ${\rm BaCo_2As_2}$. The MFT prediction for $C_{\rm mag}(T)$ with $S=7/2$ and $T_{\rm N} = 14.5$~K is shown as the black solid line. (b)~Plot of the data in~(a) as $C_{\rm mag}(T)/T$ versus~$T$, where again the MFT prediction is given by the solid black line. (c)~Magnetic entropy $S_{\rm mag}(T)$ calculated from the experimental $C_{\rm mag}(T)/T$ data in~(b) using Eq.~(\ref{Eq:SmagCalc}). The horizontal dashed line in~(c) is the theoretical high-$T$ limit  $S_{\rm mag}=R\ln(2S+1)=R\ln(8) = 17.29$~J/mol\,K for $S=7/2$.}
\label{Fig:Cmag}
\end{figure}

\begin{figure}
\includegraphics[width=3in]{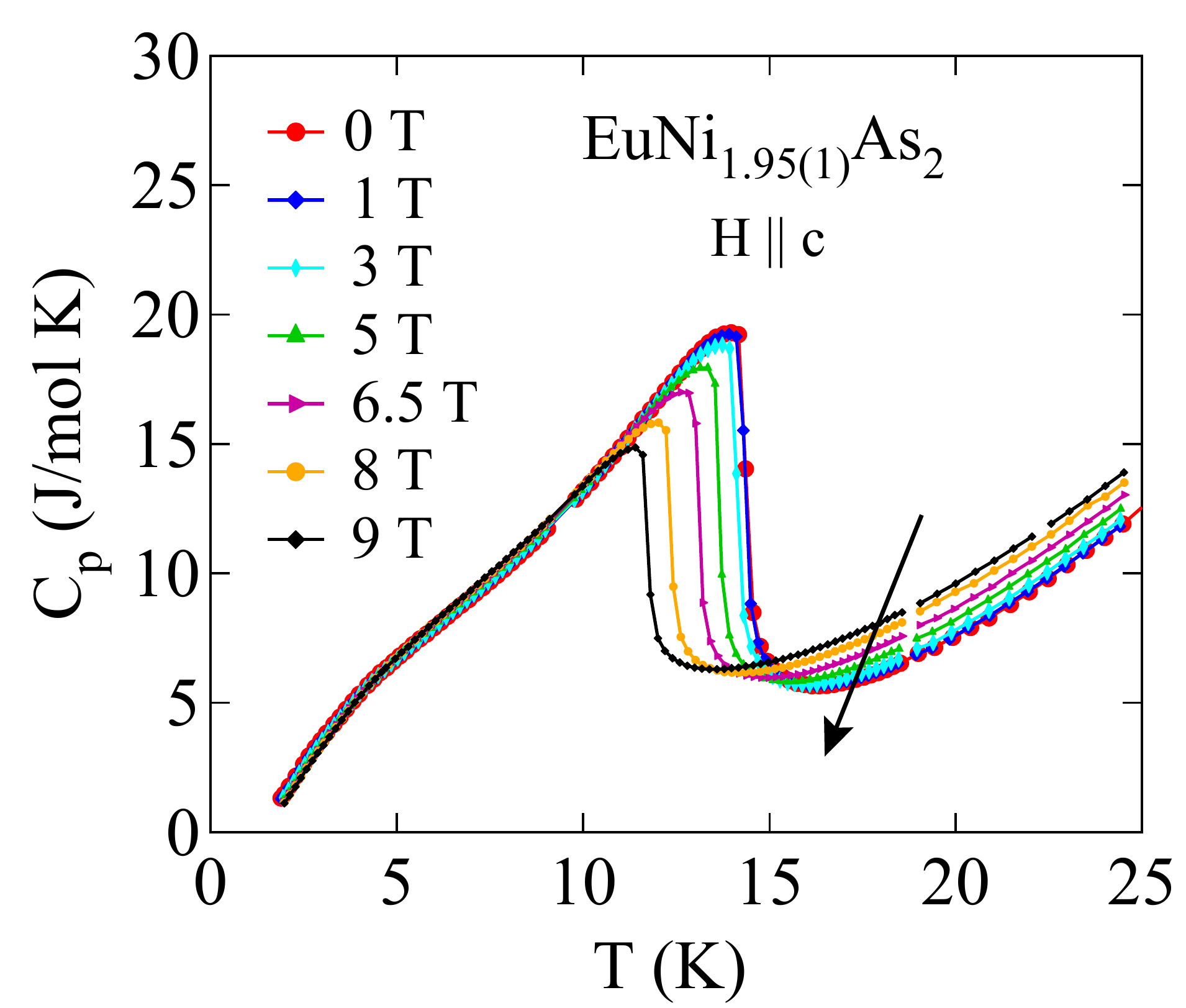}
\caption{Heat capacity $C_{\rm p}$ versus temperature~$T$ of single-crystal EuNi$_{1.95}$As$_2$ in various magnetic fields applied parallel to the $c$~axis, $H\parallel c$.}
\label{Fig:cpH}
\end{figure}

The zero-field heat capacities $C_{\rm p}(T)$ for EuNi$_{1.95}$As$_2$ and for the nonmagnetic reference compound ${\rm BaCo_2As_2}$~\cite{Sangeetha2018} measured in the temperature range from 1.8 to 300~K are shown in Fig.~\ref{Fig:cp0T}(a). A pronounced anomaly at $T_{\rm N} \approx 14.4$~K in the expanded plot in Fig.~\ref{Fig:cp0T}(b) is observed that confirms the intrinsic nature of AFM ordering in this compound. $C_{\rm p}$ attains a value of $\approx 123.7$~J/mol~K at $T=300$~K which is close to the classical Dulong-Petit high-$T$ limit value $C_{\rm V}=3nR=123.5$~J/mol~K, where $n=4.95$ is the number of atoms per formula unit and $R$ is the molar gas constant. We fitted the $C_{\rm p}(T)$ data in the PM state using the sum of an electronic term and the lattice term given by the Debye model, according to
\begin{subequations}
\label{Eq:Cpdebye}
\begin{equation}
C_{\rm p}(T) = \gamma T+ nC_{\rm V \,Debye}(T),
\end{equation}
where $\gamma$ is the Sommerfeld electronic heat capacity coefficient, $C_{\rm V \,Debye}(T)$ is the Debye lattice heat capacity given by
\begin{equation}
C_{\rm V \,Debye}(T) = 9R\left(\frac{T}{\Theta_{\rm D}}\right)^3\int_{0}^{\Theta_{\rm D}/T}\frac{x^4 dx}{(e^x-1)^2}dx,
\end{equation}
\end{subequations}
and $n=4.95$ is again the number of atoms per formula unit.  The solid curve in Fig.~\ref{Fig:cp0T}(a) represents the fit of the $C_{\rm p}(T)$ data for ${\rm 50~K\leq {\it T} \leq 300~K}$ by Eqs.~(\ref{Eq:Cpdebye}) obtained using the accurate analytic Pad\'{e} approximant function for $C_{\rm V \,Debye}$ versus  $T/\Theta_R$ given in Ref.~\cite{Goetsch2012}. The fit gave \mbox{$\gamma \sim 5$~mJ/mol~K$^2$} and $\Theta\rm_D = 284(1)$~K\@.

The magnetic contribution $C_{\rm mag}(T)$ to $C_{\rm p}(T)$ of EuNi$_{1.95}$As$_2$ obtained after subtracting the lattice contribution, taken to be $C_{\rm p}(T)$ of ${\rm BaCo_2As_2}$~\cite{Sangeetha2018}, is shown in Fig.~\ref{Fig:Cmag}(a). The MFT prediction for $C_{\rm mag}(T)$ is~\cite{Johnston2015}
\begin{equation}
\frac{C_{\rm mag}(t)}{R} = \frac{3S\bar{\mu}_0^2(t)}{(S+1)t\left[\frac{(S+1)t}{3B_S^\prime(t)} - 1\right]},
\end{equation}
which is calculated using Eqs.~(\ref{Eqs:Chixy})--(\ref{Eq:BSy}).  The solid black curve in Fig.~\ref{Fig:Cmag}(a) represents the MFT prediction for $C_{\rm mag}(T)$ calculated for $T_{\rm N} = 14.5$~K and $S=7/2$, which is seen to agree with the trend of the data.  Within MFT the discontinuity in $C_{\rm mag}$ at $T=T_{\rm N}$ is given by~\cite{Johnston2015}
\begin{eqnarray}
\Delta C{\rm_{mag}}=R\frac{5S(1+S)}{1+2S+2S^2} = 20.14~ \rm {J/mol~K},
\label{Eq:deltaCp}
\end{eqnarray}
where the second equality is calculated for $S = 7/2$.  The experimental heat capacity jump in Fig.~\ref{Fig:Cmag}(a) at $T_{\rm N} =14.2$~K is $\approx 17$~J/mol~K, somewhat smaller than the theoretical prediction. We also find that $C_{\rm mag}(T)$ is nonzero for $T{\rm_N}<T\leq 30~$K, indicating the presence of dynamic short-range AFM ordering of the Eu spins above $T_{\rm N}$, thus accounting at least in part for the decrease in $\Delta C_{\rm mag}$ from the theoretical value.

The hump in $C_{\rm mag}(T)$ below $T_{\rm N}$ at about 5~K in Fig.~\ref{Fig:Cmag}(a) is reproduced by the MFT prediction and arises naturally within MFT in the ordered-state $C_{\rm mag}(T)$ when $S$ becomes large~\cite{Johnston2015}, as shown more clearly as a plateau in the experimental and theoretical plots of $C_{\rm mag}/T$ versus~$T$ in Fig.~\ref{Fig:Cmag}(b).  This feature has also been observed in the ordered state of other spin-7/2 ${\rm Eu^{+2}}$ and ${\rm Gd^{+3}}$ compounds such as in Refs.~\cite{{Vining1983},{Bouvier1991},{Hossain2003},{Anand2015}}.  

The magnetic entropy is calculated from the $C_{\rm mag}(T)$ data for ${\rm EuNi_{1.95}As_2}$ in Fig.~\ref{Fig:Cmag}(a) using 
\begin{equation}
S_{\rm mag}(T) = \int_{0}^{T}\frac{C_{\rm mag}(T^\prime)}{T^\prime}dT^\prime
\label{Eq:SmagCalc}
\end{equation}
 and the result is shown in Fig.~\ref{Fig:Cmag}(c). The horizontal dashed line is the theoretical high-$T$ limit  $S_{\rm mag}(T) = R~{\rm ln}(2S+1) = 17.29$~J/mol~K for $S=7/2$. The entropy reaches 89$\%$ of $R{\rm ln}(8)$ at $T_{\rm N}$\@.  The high-temperature limit of the data is smaller than the theoretical value, which is likely due to a small error in estimating the lattice contribution to $C_{\rm p}(T)$.

The $C_{\rm p}(H,T)$ of single-crystal EuNi$_{1.95}$As$_2$ measured in various magnetic fields applied along the $c$~axis is shown in Fig.~\ref{Fig:cpH}. It is evident that $T_{\rm N}$ shifts to lower temperatures and the heat capacity jump at $T_{\rm N}$ decreases with increasing field, both as predicted from MFT for a field parallel to the helix axis \cite{Johnston2015}. We take $T_{\rm N}$ at each field to be the temperature of the peak in $C_{\rm p}$ versus~$T$ instead of the temperature at half-height of the transition, because the latter is ambiguous to estimate due to the significant contribution of short-range AFM ordering to $C_{\rm p}$ above $T_{\rm N}$\@.  Table~\ref{Tab:Tn} lists the $H$ dependence of $T_{\rm N}$ obtained from the $\rho(T)$ and $C_{\rm p}(T)$ measurements for $H_\perp\equiv H\parallel c$ in Figs.~\ref{Fig:ResH} and \ref{Fig:cpH}, respectively.

\begin{table}
\caption{\label{Tab:Tn} The transition temperature $T_{\rm N}$ estimated from electrical resistivity $\rho(T)$ and heat capacity $C_{\rm p}(T)$ data at various magnetic fields~$H$\@.}
\begin{ruledtabular}
\begin{tabular}{lcc}
				
$H$(T)			
& $T{\rm_N}$ from~$\rho$ 	& $T{\rm_N}$ from~$C_{\rm p}$	  \\
& (K)     		 			
& (K)  			  \\
\hline
0 	  	&14.2(8)				&13.97(19)		\\
1	 	&14.1(6)				&13.91(19)		\\
2		 &14.09(4)				&			\\
3		 &13.6(3)				&13.54(19)		\\
5		 &13.1(5)				&13.15(19)		\\
6.5		 &					&12.43(20)		\\
8		&12.3(2)				&11.82(20)		\\
9		&					&11.20(20)		\\
\end{tabular}
\end{ruledtabular}
\end{table}

\begin{figure}
\includegraphics[width=3in]{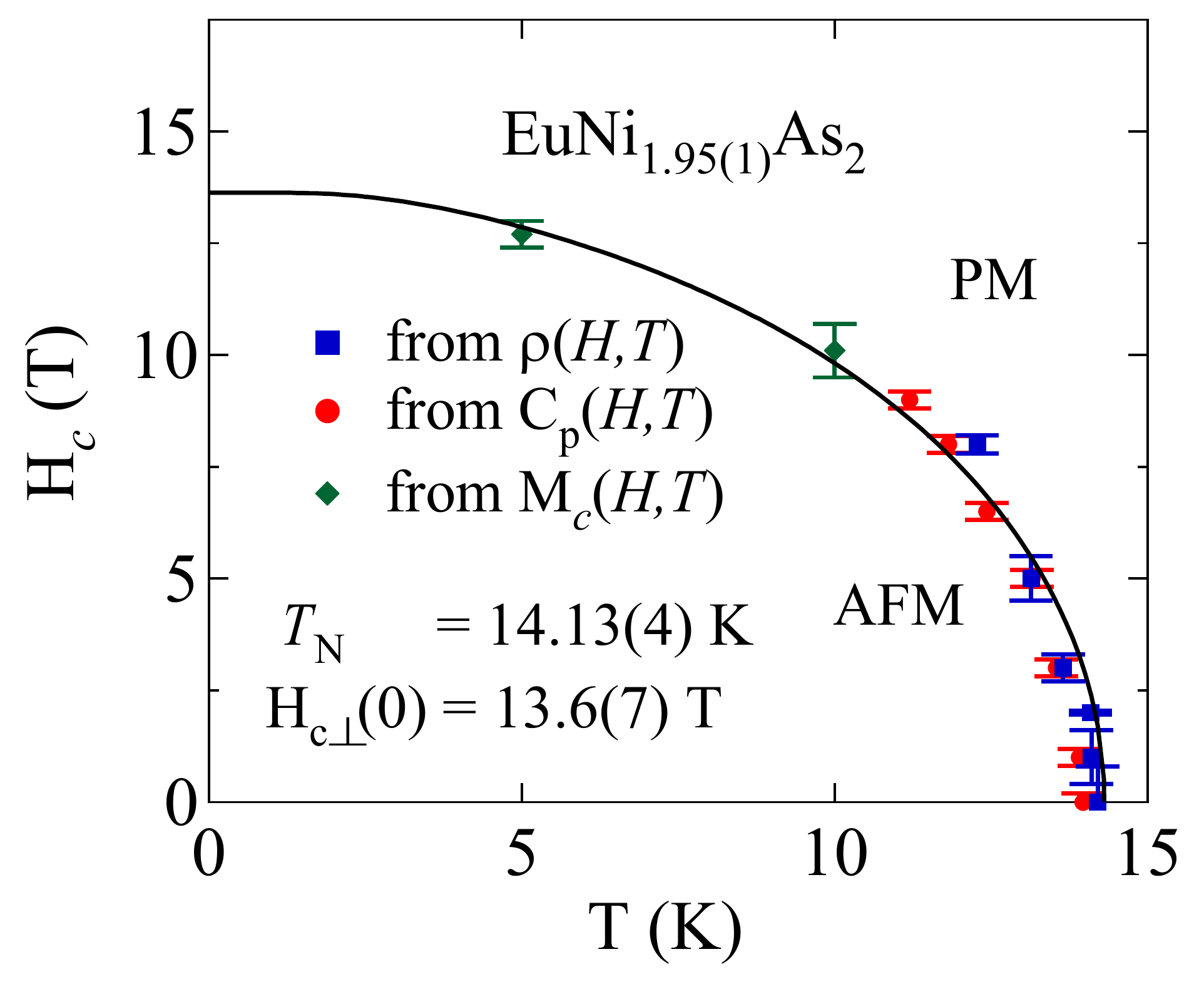}
\caption{Magnetic $H_c$-$T$ phase diagram containing the antiferromagnetic (AFM) and paramagnetic (PM) phases for single-crystal EuNi$_{1.95}$As$_2$ as determined from heat capacity $C_{\rm p}(H,T)$, magnetization $M_c(H,T)$, and electrical resistivity $\rho (H,T)$ measurements with fields $H_c$ parallel to the tetragonal $c$~axis. The solid curve is a MFT fit to the $T_{\rm N}(H,T)$ data in Table~\ref{Tab:Tn} and the $H_{\rm c}^c(T)$ data in Table~\ref{Tab:MH} by Eq.~(\ref{Eq:pd}). The values of the fitted free parameters $T_{\rm N}$ and the critical field $H_{{\rm c}\perp}(T=0)$ perpendicular to the zero-field $ab$~plane of the ordered moments are given in the figure.}
\label{Fig:PD}
\end{figure}

The $H_c$-$T$ phase diagram constructed using the data in Table~\ref{Tab:Tn} together with the $c$-axis critical field $H_{\rm c\perp}(T)$ data in Table~\ref{Tab:MH} is shown in Fig.~\ref{Fig:PD}.  According to MFT, the perpendicular critical field at which a helical AFM undergoes a second-order-transition to the PM state with increasing field at fixed $T$ is given by~\cite{Johnston2015}

\begin{equation}
H_{{\rm c}\perp}(t) = H_{\rm c\perp}(0)\bar{\mu}_0(t),
\label{Eq:pd}
\end{equation}
where $t = T/T_{\rm N}$ and the reduced $T$-dependent ordered moment $\bar{\mu}_0(t) = \mu(t)/\mu_{\rm sat} = \mu(t)/(gS\mu_{\rm B})$ is obtained by numerically solving Eq.~(\ref{Eq:barmuSoln}). The data in Fig.~\ref{Fig:PD} were fitted by Eq.~(\ref{Eq:pd}) which yielded $T_{\rm N} = 14.13(4)$~K and $H_{c\perp}(T=0)=13.6(7)$~T as shown by the solid curve in Fig.~\ref{Fig:PD}. 

The $T=0$ critical field can be independently estimated from~\cite{Johnston2015}
\begin{subequations}
\begin{equation}
\label{Eq:Hcperp}
H_{\rm c\perp}(0)=\frac{3k{\rm_B}T{\rm_N}(1-f)}{g\mu{\rm_B}(S+1)},
\end{equation}
where $f\equiv\theta_{\rm p}/T_{\rm N}$.
 Using $g=2$ and $S=7/2$ for Eu$^{+2}$, $H_{\rm c\perp}(0)$ in Eq.~(\ref{Eq:Hcperp}) can be expressed as~\cite{Johnston2015}
\begin{equation}
H_{\rm c\perp}(T=0)~({\rm T}) = 0.4962(1-f)T_{\rm N}~({\rm K}).
\label{Eq:hc}
\end{equation}
Using $T_{\rm N} = 14.13$~K obtained from the fit to the data in Fig.~\ref{Fig:PD} together with $f=-1.2(2)$ for $H=0.1$ and~1~T from Table~\ref{Tab.chidata}, Eq.~(\ref{Eq:hc}) gives
\begin{equation}
H_{c\perp}(0) \approx 16(3)~\rm T.
\label{Eq:Hcperp0Eq}
\end{equation}
\end{subequations}
This value is of the same order as the value obtained by fitting the data in the phase diagram in Fig.~\ref{Fig:PD}.

\section{\label{Sec:Summary} Summary and Conclusions}

We report the crystallographic, magnetic, thermal, and transport properties of EuNi$_2$As$_2$ single crystals grown in Bi flux. A single-crystal x-ray diffraction refinement and energy-dispersive x-ray composition analysis consistently revealed vacancies on the Ni site corresponding to the composition ${\rm EuNi_{1.95(1)}As_2}$. The occurrence of vacancies on the Ni site has not reported in previous studies of EuNi$_2$As$_2$ \cite{{Raffius1993},{Ghadraoui1988},{Jin2019}}. Reference~\cite{Jeitschko1988} found a deviation from full occupancy on the Ni site in EuNi$_2$As$_2$, but this deviation was consistent with full occupancy to within the error bar.

The electrical resistivity data indicate metallic behavior and were fitted very well in the normal state by the Bloch-Gr\"uneisen model for acoustic electron-phonon scattering.

 Our magnetic data for this helical antiferromagnet were analyzed using unified molecular field theory and yielded a helical turn angle $kd$ that is close to that found by neutron diffraction. The $M_c(H)$ data along the helix axis at 2~K are nearly linear in field up to $H=14$~T, as expected within MFT for a helical AFM structure. The perpendicular critical field $H_{c\perp}(T=0)$ for the second-order transition from the AFM to PM state is found from a fit to the data in Fig.~\ref{Fig:PD} as $H_{\rm c\perp}(T=0) = 13.6$~T\@.  We also analyzed the $\chi(T)$ data of ${\rm EuRh_2As_2}$ similar to  EuNi$_{1.95}$As$_2$ using our MFT, where a helix turn angle $kd=0.83\pi$\,rad was inferred from the fit.

The magnetic heat capacity of EuNi$_2$As$_2$ is described rather well by MFT, except for a significant amount of short-range AFM order above $T_{\rm N}$ which reduces the magnetic entropy at $T_{\rm N}$ from the value expected for spin~$S=7/2$ with $g=2$.  

The $H_c$-$T$ phase diagram for fields parallel to the helix $c$~axis was determined from the field-dependent magnetization, heat capacity, and electrical resisitivity data.  The phase line separating the AFM and PM phases is described very well by MFT.

\begin{table*}
\caption{\label{Tab.survey} Survey of the magnetic properties of ${\rm ThCr_2Si_2}$-type Eu$M_2Pn_2$ compounds with $M=$~Fe, Co, Ni, Cu, Rh and $Pn=$~P or~As.  Shown are the tetragonal $c/a$ ratio, the tetragonal crystal stucture [uncollapsed tetragonal (ucT) or collapsed tetragonal~(cT)], the magnetic transition temperature $T_{\rm m}$ which is ferromagnetic (FM) or antiferromagnetic (AFM), the AFM structure if AFM, the ordering axis or plane, and the reference for the work, where PW means present work.}
\begin{ruledtabular}
\begin{tabular}{llclcccc}	
 Compound				& $c/a$ 		& Structure	&  $T_{\rm m\,Eu}$~(K) 	& Ordering Type 	& AFM structure	 & Ordering axis/plane	& Ref. \\
\hline
${\rm EuFe_2P_2}$			&2.9433(1)	& ucT			& 29			& FM\footnotemark[1]	&  &  $c$ axis	 & \cite{{Feng2010},{Ryan2011}}	\\
${\rm EuFe_2As_2}$			&3.113(4)		& ucT			& 19			& AFM		& $A$-type\footnotemark[2]	& $ab$ plane	& \cite{{Raffius1993},{Xiao2009}}	\\
${\rm EuCo_2As_2}$			&2.930(2)		& ucT			& 47			& AFM		& Helical		& $ab$ plane	&\cite{{Raffius1993},{Tan2016}}	\\
${\rm EuCo_2P_2}$			&3.015(3)		& ucT			& 66			& AFM		& Helical 	& $ab$ plane		&\cite{{Morsen1988},{Reehuis1992}}	\\

${\rm EuCu_2As_2}$			&2.4022(1)	& cT				& 17.5		& AFM		& 		&$ab$ plane		&\cite{{Anand2015},{Sengupta2005}}	\\
${\rm EuCu_{1.75}P_2}$		&2.3312(1)	& cT				& 51			& FM		& 	&						&\cite{{Huo2011},{Mewis1980}}	\\
${\rm EuRh_2As_2}$			&2.772(3)		& ucT			& 47			& AFM		& Helical &$ab$ plane		&PW, \cite{{Singh2009},{Nandi2009}}	\\
${\rm EuRu_2As_2}$			& 2.576(1)		& cT				& 17.3		& FM		& 			& $c$ axis 			&\cite{Jiao2012}	\\
${\rm EuNi_2P_2}$\footnotemark[3]			& 2.407(1)		& cT				& 		&  	&				&  		&\cite{{Marchand1978},{Morsen1988}}	\\
${\rm EuNi_2As_2}$			& 2.448(1)		& cT 				&14			& AFM	&  Helical	& $ab$ plane		&PW, \cite{{Raffius1993},{Ghadraoui1988},{Jin2019},{Jeitschko1988}}\\ 	
\end{tabular}
\end{ruledtabular}
\footnotetext[1]{Canted at 17(3)$^\circ$ from the $c$~axis.}
\footnotetext[2]{This A-type structure with the moments aligned in the $ab$~plane is a $c$-axis helical structure with a 180$^\circ$ turn angle between adjacent moment layers along the $c$~axis.  This helical structure is explicable in terms of the $J_0$-$J_1$-$J_2$ model in Eq.~(\ref{Eq:coskdJ1J2}) if $J_1=4J_2>0$ (both AFM).}
\footnotetext[3]{Eu is intermediate valent: no magnetic ordering.}
\end{table*}

A summary of some of the properties of ${\rm ThCr_2Si_2}$-type Eu$M_2Pn_2$ compounds with $M=$~Fe, Co, Ni, Cu, Rh, or~Ru and $Pn=$~P or~As is given in Table~\ref{Tab.survey}~\cite{Raffius1993, Xiao2009, Huo2011, Jiao2012, Morsen1988, Reehuis1992, Tan2016, Anand2015, Singh2009, Ghadraoui1988, Jin2019, Jeitschko1988, Feng2010, Ryan2011, Sengupta2005, Mewis1980, Marchand1978, Nandi2009}, including the results of our measurements and analyses of EuNi$_{1.95}$As$_2$ and ${\rm EuRh_2As_2}$. One sees from Table~\ref{Tab.survey} that four of the five Eu compounds for which AFM ordering was reported have a $c$-axis helical structure, including $\rm{EuFe_2As_2}$ which has an A-type AFM structure with the Eu moments aligned in the $ab$~plane and hence is also a $c$-axis helix with a turn angle of 180$^\circ$.  The other compounds ${\rm EuFe_2P_2}$, ${\rm EuCu_{1.75}P_2}$, and ${\rm EuRu_2As_2}$ instead order ferromagnetically. Whether the structure is cT or ucT seems not to have a definitive influence on the AFM versus FM ordering of these compounds. A more general review of the correlations between the magnetic properties with respect to the cT and ucT structures of ThCr$_2$Si$_2$-structure materials is given in Sec.~VIII of Ref.~\cite{Anand2012}.  Electronic structure calculations would be of interest to determine why some compounds in Table~\ref{Tab.survey} exhibit (helical) AFM structures while the others become ferromagnetic.
 
\acknowledgments

The research at Ames Laboratory was supported by the U.S. Department of Energy, Office of Basic Energy Sciences, Division of Materials Sciences and Engineering.  Ames Laboratory is operated for the U.S. Department of Energy by Iowa State University under Contract No.~DE-AC02-07CH11358.



\begin{thebibliography}{99}

\bibitem{Just1996} G. Just and P. Paufler, On the coordination of ${\rm ThCr_2Si_2}$ (BaAl$_4$)-type compounds within the field of free parameters, J. Alloys Compd. {\bf 232}, 1 (1996).

\bibitem{Stewart1984} G. R. Stewart, Heavy-fermion systems, Rev. Mod. Phys. {\bf 56}, 755 (1984).

\bibitem{Johnston2010} D. C. Johnston, The puzzle of high temperature superconductivity in layered iron pnictides and chalcogenides, Adv. Phys. {\bf 59}, 803 (2010).

\bibitem{Canfield2010} P. C. Canfield and S. L. Bud'ko, FeAs-Based Superconductivity: A Case Study of the Effects of Transition Metal Doping on ${\rm BaFe_2As2}$, Annu. Rev. Condens. Matter Phys. {\bf 1}, 27 (2010).

\bibitem{Stewart2011} G. R. Stewart, Superconductivity in iron compounds, Rev. Mod. Phys. {\bf 83}, 1589 (2011).

\bibitem{Scalapino2012} D. J. Scalapino, A common thread: The pairing interaction for unconventional superconductors, Rev. Mod. Phys. {\bf 84}, 1383 (2012).

\bibitem{Dai2012} P. Dai, J. Hu and E. Dagotto, Magnetism and its microscopic origin in iron-based high-temperature superconductors, Nat. Phys. {\bf 8}, 709 (2012).

\bibitem{Raffius1993} H. Raffius, E. M\"{o}rsen, B. D. Mosel, W. M\"{u}ller-Warmuth, W. Jeitschko, L.Terb\"{u}chte, and T. Vomhof, Magnetic Properties Of Ternary Lanthanoid Transition Metal Arsenides Studied by M\"{o}ssbauer and Susceptibility Measurements, J. Phys. Chem. Solids \textbf{54}, 135 (1993).

\bibitem{Ren2008} Z. Ren, Z. W. Zhu, S. Jiang, X. F. Xu, Q. Tao, C. Wang, C. M. Feng, G. H. Cao, and Z. A. Xu, Antiferromagnetic transition in ${\rm EuFe_2As_2}$: A possible parent compound for superconductors, Phys. Rev. B {\bf 78}, 052501 (2008).

\bibitem{Tegel2008} M. Tegel, M. Rotter, V. Wei\ss, F. M. Schappacher, R. P\"ottgen and D. Johrendt, Structural and magnetic phase transitions in the ternary iron arsenides ${\rm SrFe_2As_2}$ and ${\rm EuFe_2As_2}$,  J. Phys: Condens. Matter {\bf 20}, 452201 (2008).

\bibitem{Xiao2009}Y. Xiao, Y. Su, M. Meven, R. Mittal, C. M. N. Kumar, T. Chatterji, S. Price, J. Persson, N. Kumar, S. K. Dhar,
A. Thamizhavel, and Th. Brueckel, Magnetic structure of ${\rm EuFe_2As_2}$ determined by single-crystal neutron diffraction, Phys. Rev. B {\bf 80}, 174424 (2009).

\bibitem{Terashima2009}T. Terashima, M. Kimata, H. Satsukawa, A. Harada, K. Hazama, S. Uji, H. S. Suzuki, T. Matsumoto, and K. Muruta, ${\rm EuFe_2As_2}$ under High pressure: An Antiferromagnetic Bulk Superconductor, J. Phys. Soc. Jpn. {\bf 78}, 083701 (2009).

\bibitem{Zapf2017} S. Zapf and M. Dressel, Europium-based iron pnictides: a unique laboratory for magnetism, superconductivity and structural effects, Rep. Prog. Phys. {\bf 80}, 016501 (2017).

\bibitem{Huo2011} D. Huo, J. Lin, G. Tang, L. Li, Z. Qian, and T. Takabatake, Ferromagnetic ordering in ${\rm EuCu_2P_2}$, J. Phys.: Conf. Ser. {\bf 263}, 012014 (2011).

\bibitem{Jiao2012} W. H. Jiao, I. Felner, I. Nowik, and G. H. Cao, ${\rm EuRu_2As_2}$: A New Ferromagnetic Metal with Collapsed ${\rm ThCr_2Si_2}$-Type Structure, J. Supercond. Nov. Magn. {\bf 25}, 441 (2012).


\bibitem{Morsen1988} E. M\"{o}rsen, B. D. Mosel, W. M\"{u}ller-Warmuth, M. Reehuis, and W. Jeitschko, M\"{o}ssbauer and Magnetic Susceptibility Investigations of Strontium, Lanthanum and Europium Transition Metal Phosphides with ${\rm ThCr_2Si_2}$ Type Structure, J. Phys. Chem. Solids \textbf{49}, 785 (1988).

\bibitem{Reehuis1992}  M. Reehuis, W. Jeitschko, M. H. M\"{o}ller, and P. J. Brown, A Neutron Diffraction Study of the Magnetic Structure of ${\rm EuCo_2P_2}$, J. Phys. Chem. Solids {\bf 53}, 687 (1992).

\bibitem{Chefki1998} M. Chefki, M. M. Abd-Elmeguid, H. Micklitz, C. Huhnt, W. Schlabitz, M. Reehuis, and W. Jeitschko, Pressure-induced Transition of the Sublattice Magnetization in ${\rm EuCo_2P_2}$: Change from Local Moment Eu (4$f$) to Itinerant Co (3$d$) Magnetism, Phys. Rev. Lett. {\bf 80}, 802 (1998).

\bibitem{Sangeetha2016} N. S. Sangeetha, E. Cuervo-Reyes, A. Pandey, and D. C. Johnston, ${\rm EuCo_2P_2}$: A model molecular-field helical Heisenberg antiferromagnet, Phys. Rev. B {\bf 94}, 014422 (2016).

\bibitem{Marchand1978} R. Marchand and W. Jeitschko, Ternary Lanthanoid--Transition Metal Pnictides with ${\rm ThCr_2Si_2}$-Type Structure, J. Solid State Chem. \textbf{24}, 351 (1978).

\bibitem{Tan2016} X. Tan, G. Fabbris, D. Haskel, A. A. Yaroslavtsev, H. Cao, C. M. Thompson, K. Kovnir, A. P. Menushenkov, R. V. Chernikov, V. O. Garlea, and M. Shatruk, A Transition from Localized to Strongly Correlated Electron Behavior and Mixed Valence Driven by Physical or Chemical Pressure in ${\rm ACo_2As_2}$ (A = Eu and Ca), J. Am. Chem. Soc. {\bf 138}, 2724 (2016).

\bibitem{Sangeetha2018}  N. S. Sangeetha, V. K. Anand,  E.  Cuervo-Reyes,  V. Smetana, A.-V. Mudring, and D. C. Johnston, Enhanced moments of Eu in single crystals of the metallic helical antiferromagnet EuCo$_{2-y}$As$_2$, Phys. Rev. B {\bf 97}, 144403 (2018).

\bibitem{Bishop2010} M. Bishop, W. Uhoya, G. Tsoi, Y. K. Vohra, A. S. Sefat, and B. C. Sales, Formation of collapsed tetragonal phase in ${\rm EuCo_2As_2}$ under high pressure, J. Phys.: Condens. Matter {\bf 22}, 425701 (2010).

\bibitem{Anand2015} V. K. Anand and D. C. Johnston, Antiferromagnetism in ${\rm EuCu_2As_2}$ and ${\rm EuCu_{1.82}Sb_2}$ single crystals, Phys. Rev. B {\bf 91}, 184403 (2015). 

\bibitem{Singh2009} Y. Singh, Y. Lee, B. N. Harmon, and D. C. Johnston, Unusual magnetic, thermal, and transport behavior of single crystalline ${\rm EuRh_2As_2}$, Phys. Rev. B {\bf 79}, 220401(R) (2009).

\bibitem{Bauer2008} E. D. Bauer, F. Ronning, B. L. Scott,  and J. D. Thompson, Superconductivity in ${\rm SrNi_2As_2}$ single crystals, Phys. Rev. B {\bf 78}, 172504 (2008).

\bibitem{Ronning2009} F. Ronning, E. D. Bauer, T. Park, S.-H. Baek, H. Sakai, and J. D. Thompson, Superconductivity and the effects of pressure and structure in single-crystalline ${\rm SrNi_2P_2}$, Phys. Rev. B {\bf 79}, 134507 (2009).

\bibitem{Ronning2008} F. Ronning, N. Kurita,  E. D. Bauer, B. L. Scott, T. Park,T. Klimczuk, R. Movshovich and J. D. Thompson, The first order phase transition and superconductivity in ${\rm BaNi_2As_2}$ single crystals, J. Phys.: Condens. Matter {\bf 20}, 342203 (2008).

\bibitem{Mine2008} T. Mine, H. Yanagi, T. Kamiya, Y. Kamihara, M. Hirano, and H. Hosono, Nickel-based phosphide superconductor with infinite-layer structure, ${\rm BaNi_2P_2}$, Solid State Commun. {\bf 147}, 111 (2008).

\bibitem{Jeevan2008} H. S. Jeevan, Z. Hossain, D. Kasinathan, H. Rosner, C. Geibel, and P. Gegenwart, High-temperature superconductivity in ${\rm Eu_{0.5}K_{0.5}Fe_2As_2}$, Phys. Rev. B {\bf 78}, 092406 (2008).

\bibitem{Maiwald2012} J. Maiwald, H. S. Jeevan, and P. Gegenwart, Signatures of quantum criticality in hole-doped and chemically pressurized ${\rm EuFe_2As_2}$ single crystals, Phys. Rev. B {\bf 85}, 024511 (2012).

\bibitem{Jiao2012a} W. H. Jiao, J. K Bao, Q. Tao, H. Jiang, C. M Feng, Z. A Xu, and G. H. Cao, Evolution of superconductivity and ferromagnetism in ${\rm Eu(Fe_{1-\it x}Ru_{\it x})_2As_2}$, J. Phys.: Conf. Ser. {\bf 400}, 022038 (2012).

\bibitem{Jiang2009} S. Jiang, H. Xing, G. Xuan, Z. Ren, C. Wang, Z. A. Xu, and G. H. Cao, Superconductivity and local-moment magnetism in ${\rm Eu(Fe_{0.89}Co_{0.11})_2As_2}$, Phys. Rev. B {\bf 80}, 184514 (2009).

\bibitem{Paramanik2013} U. B. Paramanik, D. Das, R. Prasad, and Z. Hossain, Reentrant superconductivity in ${\rm Eu(Fe_{1-\it x}Ir_{\it x})_2As_2}$, J. Phys.: Condens. Matter {\bf 25}, 265701 (2013).

\bibitem{Ren2009} Z. Ren, Q. Tao, S. Jiang, C. Feng, C. Wang, J. Dai, G. H. Cao, and Z. A. Xu, Superconductivity Induced by Phosphorus Doping and its Coexistence with Ferromagnetism in ${\rm EuFe_2(As_{0.7}P_{0.3})_2}$, Phys. Rev. Lett. {\bf 102}, 137002 (2009).

\bibitem{Jeevan2011} H. S. Jeevan, D. Kasinathan, H. Rosner, and P. Gegenwart, Interplay of antiferromagnetism, ferromagnetism, and superconductivity in ${\rm EuFe_2(As_{1-\it x}P_{\it x})_2}$ single crystals, Phys. Rev. B {\bf 83}, 054511 (2011).

\bibitem{Li2009} L. J. Li, Y. K. Luo, Q. B. Wang, H. Chen, Z. Ren, Q. Tao, Y. K. Li, X. Lin, M. He, Z. W. Zhu, G. H. Cao, and Z. A. Xu, Superconductivity induced by Ni doping in ${\rm BaFe_2As_2}$ single crystals, New J. Phys. {\bf 11}, 025008 (2009).

\bibitem{Nowik2011} I. Nowik, I. Felner, Z. Ren, G. H. Cao, and Z. A. Xu, ${\rm ^{57}Fe}$ and  ${\rm ^{151}Eu}$ M\"{o}ssbauer spectroscopy and magnetization studies of ${\rm Eu(Fe_{0.89}Co_{0.11})_2As_2}$ and ${\rm Eu(Fe_{0.9}Ni_{0.1})_2As_2}$, New J. Phys. {\bf 13}, 023033 (2011).

\bibitem{Ren2009a} Z. Ren, X. Lin, Q. Tao, S. Jiang, Z. Zhu, C. Wang, G. Cao, and Z. Xu, Suppression of spin-density-wave transition and emergence of ferromagnetic ordering of ${\rm Eu^{+2}}$ moments in ${\rm EuFe_{2-\it x}Ni_{\it x}As_2}$, Phys. Rev. B {\bf 79}, 094426 (2009).

\bibitem{Ghadraoui1988} E. H. El Ghadraoui, J. Y. Pivan, R. Gu\'{e}rin, O. Pena, J. Padiou, and M. Sergent, Polymorphism and Physical properties of ${\rm LnNi_2As_2}$ compounds (Ln = La$\rightarrow$Gd), Mater. Res. Bull. {\bf 23}, 1345 (1988).

\bibitem{Jin2019} W. T. Jin, N. Qureshi, Z. Bukowski, Y. Xiao, S. Nandi, M. Babij, Z. Fu, Y. Su, and Th. Br\"{u}ckel, Spiral magnetic ordering of the Eu moments in  EuNi$_2$As$_2$, Phys. Rev. B {\bf 99}, 014425 (2019).

\bibitem{Johnston2012} D. C. Johnston, Magnetic Susceptibility of Collinear and Noncollinear Heisenberg Antiferromagnets, Phys. Rev. Lett. {\bf 109}, 077201 (2012).

\bibitem{Johnston2015} D. C. Johnston,  Unified molecular field theory for collinear and noncollinear Heisenberg antiferromagnets, Phys. Rev. B {\bf 91}, 064427 (2015).

\bibitem{APEX2015} APEX3, Bruker AXS Inc., Madison, Wisconsin, USA, 2015.

\bibitem{SAINT2015} SAINT, Bruker AXS Inc., Madison, Wisconsin, USA, 2015.

\bibitem{Krause2015} L. Krause, R. Herbst-Irmer, G. M. Sheldrick, and D. J. Stalke, Comparison of silver and molybdenum microfocus X-ray sources for single-crystal structure determination, J. Appl. Crystallogr. {\bf 48}, 3 (2015).

\bibitem{Sheldrick2015A} G. M. Sheldrick, SHELTX--Integrated space-group and crystal-structure determination, Acta Crystallogr. A {\bf 71}, 3 (2015).

\bibitem{Sheldrick2015C} G. M. Sheldrick, Crystal structure refinement with SHELXL, Acta Crystallogr.~C {\bf 71}, 3 (2015).

\bibitem{Jeitschko1988} W. Jeitschko, W. K. Hofmann, and L. J. Terb\"{u}chte, Lanthanoid and Uranium Nickel Arsenides with ${\rm CaBe_2Ge_2}$- and ${\rm ThCr_2Si_2}$-Type Structures, J. Less-Common Met. {\bf 137}, 133 (1988).

\bibitem{Johnston2016} D. C. Johnston, Magnetic dipole interactions in crystals, Phys. Rev. B {\bf 93}, 014421 (2016).

\bibitem{Johnston2017} D. C. Johnston, Influence of uniaxial single-Ion anisotropy on the magnetic and thermal properties of Heisenberg antiferromagnets within unified molecular field theory, Phys. Rev. B {\bf 95}, 224401 (2017).

\bibitem{Johnston2017b} D. C. Johnston, Magnetic structure and magnetization of helical antiferromagnets in high magnetic fields perpedicular to the helix axis at zero temperature, Phys. Rev. B {\bf 96}, 104405 (2017).

\bibitem{Johnston2019} D. C. Johnston, Magnetic structure and magnetization of $z$-axis helical Heisenberg antiferromagnets
with XY anisotropy in high magnetic fields transverse to the helix axis at zero temperature, Phys. Rev. B {\bf 99}, 214438 (2019).

\bibitem{Nagamiya1967} T. Nagamiya, Helical Spin Ordering---1 Theory of Helical Spin Configurations,  in {\it Solid State Physics}, Vol.~20, ed.\ F. Seitz, D. Turnbull, and H. Ehrenreich (Academic Press, New York, 1967), pp.~305--411.

\bibitem{Nandi2009} S. Nandi, A. Kreyssig, Y. Lee, Y. Singh, J. W. Kim, D. C. Johnston, B. N. Harmon, and A. I. Goldman, Magnetic ordering in ${\rm EuRh_2As_2}$ studied by x-ray resonant magnetic scattering,  Phys. Rev. B {\bf 79}, 100407(R) (2009).

\bibitem{Goetsch2012}  R. J. Goetsch, V. K. Anand, A. Pandey, and D. C. Johnston, Structural, thermal, magnetic, and electronic transport properties of the LaNi$_2$(Ge$_{1-x}$P$_x$)$_2$ system, Phys. Rev. B {\bf 85}, 054517 (2012).

\bibitem{Vining1983} C. B. Vining and R. N. Shelton, Low-temperature heat capacity of antiferromagnetic ternary rare-earth iron silicides $M_2{\rm Fe_3Si_5}$, Phys. Rev. B {\bf 28}, 2732 (1983).

\bibitem{Bouvier1991} M. Bouvier, P. Lethuillier, and D. Schmitt, Specific heat in some gadolinium compounds.~I.~Experimental, Phys. Rev. B {\bf 43}, 13137 (1991).

\bibitem{Hossain2003} Z. Hossain and C. Geibel, Magnetic properties of single crystal ${\rm EuCo_2Ge_2}$, J. Magn. Magn. Mater. {\bf 264}, 142 (2003).

\bibitem{Johnston2011}  D. C. Johnston, R. J. McQueeney, B. Lake, A. Honecker, M. E. Zhitomirsky, R. Nath, Y. Furukawa, V. P. Antropov, and Y. Singh, Magnetic exchange interactions in BaMn$_2$As$_2$: A case study of the $J_1$-$J_2$-$J_c$ Heisenberg model, Phys. Rev. B {\bf 84}, 094445 (2011).

\bibitem{Feng2010} C. Feng, Z. Ren, S. Xu, S. Jiang, Z. A. Xu, G. Cao, I. Nowik, I. Felner, K. Matsubayashi, and Y. Uwatoko, Magnetic ordering and dense Kondo behavior in ${\rm EuFe_2P_2}$,  Phys. Rev. B {\bf  82}, 094426 (2010).

\bibitem{Ryan2011} D. H. Ryan, J. M. Cardogan, Z. A. Xu, and G. H. Cao, Magnetic structure of ${\rm EuFe_2P_2}$ studied by neutron powder diffraction, Phys. Rev. B {\bf 83}, 132403 (2011).

\bibitem{Sengupta2005} K. Sengupta, P. L. Paulose, E. V. Sampathkumaran, Th. Doert, and J. P. F. Jemetio, Magnetic behavior of ${\rm EuCu_2As_2}$: A delicate balance between antiferromagnetic and ferromagnetic order, Phys. Rev. B {\bf 72}, 184424 (2005).

\bibitem{Mewis1980} A. Mewis, Ternary Phosphides with the ${\rm ThCr_2Si_2}$ Structure, Z. Naturforsch. {\bf 35b}, 141 (1980).

\bibitem{Anand2012} V. K. Anand. P. K. Perera, A. Pandey. R. J. Goetsch, A. Kreyssig, and D. C. Johnston, Crystal growth and physical properties of SrCu$_2$As$_2$, SrCu$_2$Sb$_2$, and BaCu$_2$Sb$_2$, Phys. Rev. B {\bf 85}, 214523 (2012).

\end{thebibliography}
\end{document}